\documentclass[sigconf,nonacm,prologue,table,xcdraw]{acmart}

\AtBeginDocument{%
 }

\usepackage{booktabs} 
\usepackage{color}
\usepackage{graphicx}
\usepackage{multirow}
\usepackage{bbm}
\usepackage{subcaption}
\usepackage{enumitem}
\usepackage{balance}

\begin{document}

\title{Dataset and Models for Item Recommendation Using Multi-Modal User Interactions}


\author{Simone Borg Bruun}
\orcid{0000-0003-1619-4076}
\affiliation{%
  \institution{University of Copenhagen}
  \city{Copenhagen}
  \country{Denmark}
}
\email{simoneborgbruun@di.ku.dk}

\author{Krisztian Balog}
\orcid{0000-0003-2762-721X}
\affiliation{%
  \institution{University of Stavanger}
  \city{Stavanger}
  \country{Norway}
}
\email{krisztian.balog@uis.no}

\author{Maria Maistro}
\orcid{0000-0002-7001-4817}
\affiliation{%
  \institution{University of Copenhagen}
  \city{Copenhagen}
  \country{Denmark}
}
\email{mm@di.ku.dk}

\renewcommand{\shortauthors}{Simone Borg Bruun, Krisztian Balog, \& Maria Maistro}

\begin{abstract}
While recommender systems with multi-modal item representations (image, audio, and text), have been widely explored, learning recommendations from multi-modal user interactions (e.g., clicks and speech) remains an open problem.
We study the case of multi-modal user interactions in a setting where users engage with a service provider through multiple channels (website and call center). In such cases, incomplete modalities naturally occur, since not all users interact through all the available channels.
To address these challenges, we publish a real-world dataset that allows progress in this under-researched area.
We further present and benchmark various methods for leveraging multi-modal user interactions for item recommendations, and propose a novel approach that specifically deals with missing modalities by mapping user interactions to a common feature space.
Our analysis reveals important interactions between the different modalities and that a frequently occurring modality can enhance learning from a less frequent one.

\end{abstract}





\maketitle

\section{Introduction}
\label{sec:intro}
Recommender systems (RSs) most often learn from the users’ actions such as ratings or purchases of items. In many domains, users interact with the system through multiple channels like website and call center, social tagging~\cite{Dong:2022:JISYS}, image posting~\cite{Kurt:2017:UBMK} and location sharing~\cite{Bao:2017:GIS}.
We refer to these different interaction types (e.g., clicks and speech) as multi-modal user interactions. We study the generation of recommendations when the user interactions have different modalities and thereby cannot simply be combined and used in existing recommender methods that are designed for uni-modal user interactions. 
Previous work on multi-modal RSs exclusively focuses on multi-modal representations of items such as image, audio, and text \cite{He:2016:AAAI, Wei:2019:MM, Sun:2020:CIKM, Liu:2019:MM}. These methods are not designed for our scenario with multi-modal user interactions. In particular, existing methods only work when all modality information is available during training and inference. This is a problem when dealing with multi-modal user interactions where incomplete modalities naturally occur, since not all users interact through all the available channels. For example, in the insurance domain, where items are rather complex, some users purchase items through the website, while others prefer to make the purchase over the phone. Note that our scenario distinguishes from previous work on multi-behavior RSs \cite{Xia:2021:ICDE, Xu:2023:SIGIR, Li:2023:ApplSci}, that try to infer user preferences with feedback of different categories, such as view, add-to-cart and purchase, but do not learn from user interactions of different modalities.

While multi-modal user interactions have great potential to be utilized in RSs, the lack of public datasets is a major roadblock to progress in this area.
Therefore, the primary contribution of this paper is the creation and release of a real-world dataset to facilitate RSs research on multi-modal user interactions.
The data is collected from a company dealing with insurance products for individuals and consists of (1) user sessions logged from the company’s website, (2) transcribed conversations between users and the company's insurance agent, and (3) purchase actions. This data opens up novel research opportunities to predict product purchases based on rich, multi-modal interactions.
Unlike commonly studied domains like movie, restaurant, or book recommendations, which are considered low-risk scenarios, the purchase of insurance products is a high-stakes domain where decisions can have a long-lasting impact on an individual's life, marking a significant departure from traditional RSs research.
As our second contribution, we present and experimentally compare several approaches for combining different modalities for recommendation. We consider existing methods, such as different imputation approaches \citep{Cai:2018:KDD, Tran:2017:CVPR, Wang:2018:EMNLP} and knowledge distillation \citep{Wang:2020:KDD}. We also propose a novel approach to jointly model different modalities, that suffer from naturally induced incompleteness, by mapping them into a common feature space.
Specifically, we explore the following research questions.
\begin{itemize}
    \item[RQ1] How can we effectively learn recommendations from multi-modal user interactions?
    \item[RQ2] How does it affect the quality of recommendations to jointly model the multiple modalities compared to separately?
\end{itemize}
A crucial factor for RQ1 is how we can represent multi-modal user interactions so they can be combined. In RQ2 it is essential to investigate if there are important interactions between the modalities and whether information from one modality can be useful when learning recommendations from another modality.
Experimental results show that the two modalities represent different information that supplement each other well in the recommendation task. Compared to the existing methods, our proposed approach manages to capture important interactions between the modalities as well as use information from the most frequent modality when learning recommendations from the less frequent modality.



In summary, this paper makes the following contributions: 
(1) we create and release a dataset of multi-modal user interactions for the recommendation of financial products;
(2) we present and experimentally compare various approaches for leveraging multi-modal user interactions; and
(3) we conduct an in-depth analysis of the results to shed light on what makes this problem challenging. 
All resources developed within this study, including the dataset and implementations of the models, are made publicly available on GitHub\footnote{\url{https://github.com/simonebbruun/RS_multi_modal_user_interactions}} and Zenodo.\footnote{\url{https://doi.org/10.5281/zenodo.10952736}}
\section{Related work}
\label{sec:related}
We present related work focusing on multi-modal recommendation datasets, RSs for the insurance domain, conversation-based recommendations, and multi-modal recommendation models.

\paragraph{\textbf{Multi-modal Recommendation Datasets}}
We contribute a novel dataset for multi-modal recommender tasks which is different from publicly available ones as follows.
First, our dataset contains multi-modal user interactions while existing datasets contain multi-modal item representations, e.g., visual and textual representations of movies, books, and music~\citep{Zhu:2021:dataset}, video and textual representations of micro-videos~\citep{Fei:2020:dataset}, and acoustic and textual representations of music~\citep{Thierry:2011:ISMIR}.
Second, in our dataset missing modalities naturally occur since users might not interact through all channels, while in~\citep{Zhu:2021:dataset,Fei:2020:dataset,Thierry:2011:ISMIR} missing modalities only occur due to technical reasons.

\paragraph{\textbf{Insurance Domain}}
In this domain, user feedback on items is sparse, because of few different items and because users rarely interact with insurance products.
Most prior work supplements the small volume of user feedback with user demographics, such as age, income level, and employment.
In this case, different techniques are used to categorize users based on demographic characteristics, then make recommendations within these categories \citep{Xu:2014:JSSI, Mitra:2014:IJECS, Qazi:2017:RecSys}.
\citet{Bi:2020:SIGIR} propose a cross-domain RS for the insurance domain. They use knowledge from an e-commerce source domain with daily necessities to learn better recommendations in the insurance target domain when data is sparse.
In \cite{Bruun:2022:RecSys} and \cite{Bruun:2023:TORS} a session-based approach is presented that uses a recurrent neural network (RNN) to learn insurance recommendations from web sessions with several types of user actions.
None of the above methods use conversations or multiple modalities to learn insurance recommendations.

\paragraph{\textbf{Conversation-based Recommendations}}
The way we utilize conversations is not to be confused with conversational RSs, which are interactive systems that allow the user to disclose preferences, ask questions about items, and provide feedback \citep{Jannach:2021:CSUR}. Instead, we learn recommendations from past observed conversations and make recommendations in one-shot interactions.
Few studies have focused on RSs based on past conversations.
\citet{Gentile:2011:ISWC} exploit e-mail conversations to learn user profiles utilizing different techniques (keyword extraction, extraction of named entities, and concept extraction). The profiles are then used to estimate the similarity between users.
\citet{Rosa:2019:IEEE} analyze text messages posted on social networks with the purpose of detecting users with potential psychological disorders (depression and stress). Then, if needed, an RS is used to send messages of happiness, calm, relaxation, or motivation. Text sentences are analyzed employing neural network approaches.
\citet{Torbati:2021:ECIR} explore past online chats between users to learn search-based recommendations. They leverage language modeling techniques such as entity detection and entity-based expansions to represent the chats that are then used for ranking.
However, none of these works combine the conversations with any other modalities.

\paragraph{\textbf{Multi-modal Recommendation Models}}
Multi-modal RSs can model relationships between different modalities and possibly benefit from the complementary and diverse sources of information that can not be captured by a uni-modal RS. 
\citet{He:2016:AAAI} address cold start and data sparsity with matrix factorization on the user-item matrix complemented with visual representations.
\citet{Wei:2019:MM} use graph neural networks to learn the representation of each modality and then fuse them together as the final item representations. Instead of fusing the modality representations, \citet{Sun:2020:CIKM} use knowledge graphs to include side information of different modalities. \citet{Liu:2019:MM} introduce an attention neural network when fusing visual and textual representations, which can recognize users’ varied preferences. 
All the above methods focus on multi-modal item representations, while we address the distinct challenge of multi-modal user interactions across different channels.  Also, these methods rely on complete modality information during training and inference, which is unrealistic for multi-modal user interactions where not all users engage through all channels.

To deal with missing modalities, some imputation methods have been proposed in the 
broader field of machine learning. \citet{Tran:2017:CVPR} concatenate all the modalities to form a large matrix, then apply a cascaded residual autoencoder to impute the missing elements. \citet{Cai:2018:KDD} use adversarial learning to complete the missing modalities, and \citet{Wang:2018:EMNLP} use a generative model to reconstruct the modality-specific embedding. In all these scenarios, the missing modalities existed in the real world but were missing in the dataset for technical reasons, such as problems with the measurement/tracking, or synthetically generated by removing observations from the dataset. This is a major difference from our scenario, where missing modalities are a result of ``normal use.'' For instance, if a user did not interact with the website, then it might not be the most appropriate method to generate web sessions that did not take place. Furthermore, there might be some information in the user choosing not to have a conversation or web session, which is lost when using imputation.
\citet{Wang:2020:KDD} use knowledge distillation to avoid imputation: they first train teacher models on each modality independently, then the student models are trained on complete modalities with soft labels from the teacher models and the true label.
This method does not work well when only a small amount of the samples have complete modalities, as in our case.

\section{Dataset}
\label{sec:dataset}
To the best of our knowledge, there exists no publicly available dataset that contains user interactions of different modalities and with a naturally induced incompleteness of modalities.  We use a real-life dataset that we have obtained from a commercial insurance vendor and that we make freely available to the research community.
Next, we describe the application context as well as the dataset.

\subsection{Application Context}
The data is collected from a vendor that deals with insurance for individuals. The items to be recommended are insurance products such as car, house and accident insurance or additional coverages, like roadside assistance for car insurance or chewing injury for accident insurance. The objective of an RS in this domain is not to help customers discover items in a large item space, rather it is to remind customers continuously to adjust their insurances to suit their needs.

The company's website is divided into four sections: (1) e-com\-merce, where users can buy insurance products, (2) claims reporting, where users can report insurance claims, (3) information section with information about payment methods, contact details, etc., and (4) personal account, where users can log in.
In that way, users can interact with items, like buying a \emph{car insurance} or reporting a \emph{house insurance} claim, and can interact with services, like finding information about the \emph{payment methods} or changing personal information such as \emph{address}.

The company's call center has only inbound calls (i.e., the user calling the company). Here, the user can call for exactly the same purposes as on the website; for example, to buy insurances, report claims, ask about payment methods, and change address.

\subsection{Dataset Description}
The dataset was collected between May 1, 2022, to April 30, 2023.
We collected \emph{purchases} of insurance products and additional coverages made by existing customers. A purchase consists of one or more items bought by the same user at the same time. Both purchases made on the website and over the phone are included.
For each user in the dataset, we collected all \emph{conversations} that occurred before the user's purchase. A conversation consists of transcribed sentences from phone calls between a user and an insurance agent.
The sentences in the dataset come with the order of the sentences, the speaker (user or agent), and the transcribed text.
For each user in the dataset, we moreover collected all \emph{web sessions} that occurred before the user's purchase. 
A web session consists of user actions on the insurance website. 
The actions in the dataset come with the order they are performed and action tags describing the section of the website in which the user interacts, the object on the website that a user chooses to interact with, and the way that a user interacts with objects.

All data 
has been anonymized to protect the identities and personal information of the individuals involved.
Personal identifiers such as names, addresses, and contact information were removed; demographic information is not included.
This ensures that individual participants cannot be identified or singled out from the dataset.
The anonymization of data was carried out in compliance with applicable data protection laws and regulations.
We thus expect this data to be useful for a long time without requiring frequent updates.
We might release updated text embeddings as needed to accommodate the latest advancements in language models.

\subsection{Dataset Pre-processing}
The dataset is pre-processed in the following way.
All items with frequency of less than $1\%$ are removed from the purchases since low-frequency items are not optimal for modeling.
Consecutive repeated actions of the same kind are discarded in web sessions because they very likely represent noise (e.g., double click due to latency).
All conversations with less than four sentences and all web sessions with less than three actions are removed, as they offer limited insights. 
To avoid slow processing, all conversations and web sessions are truncated in the end to have a maximum of $541$ sentences and $40$ actions respectively (the $99$th percentiles). 
Not all historical conversations and web sessions are kept for each user as only recent conversations and web sessions are assumed to be relevant to the current task. An inactivity threshold is used to define recent events (i.e., conversations/web sessions) such that two events belong to the same task if the time duration between them does not exceed a specific threshold, which is set to be $14$ days based on~\citep{Bruun:2022:RecSys}. Conversations and web sessions that exceed this threshold are thereby discarded. In addition within the 14 days rule,
the list of recent events for each user is truncated to a maximum of 10
events (the 99th percentile) to reduce training time.
Text embeddings are generated for each sentence in the conversations using a pre-trained language-specific BERT model\footnote{\url{https://huggingface.co/Maltehb/danish-bert-botxo}} 
on the raw text.
Keywords are extracted from the sentences by removing stop words, 
lemmatization, and using part-of-speech tagging to identify nouns.

\begin{table}[tb]
\centering
\caption{Main properties of the dataset (*mean/std).}
\label{tab:data_statistics}
\vspace*{-0.5\baselineskip}
\resizebox{0.5\columnwidth}{!}{%
\begin{tabular}{lr}
\hline
Users & 51,877 \\
Items & 24 \\
Purchases & 62,401 \\
Conversations & 25,515 \\
Web sessions & 115,045 \\ \hline
Conversations before purchase* & 1.38/0.82 \\
Web sessions before purchase* & 2.3/1.98 \\ \hline
\end{tabular}%
}
\end{table}

\begin{figure}[tb]
    \centering
    \includegraphics[width=0.4\columnwidth]{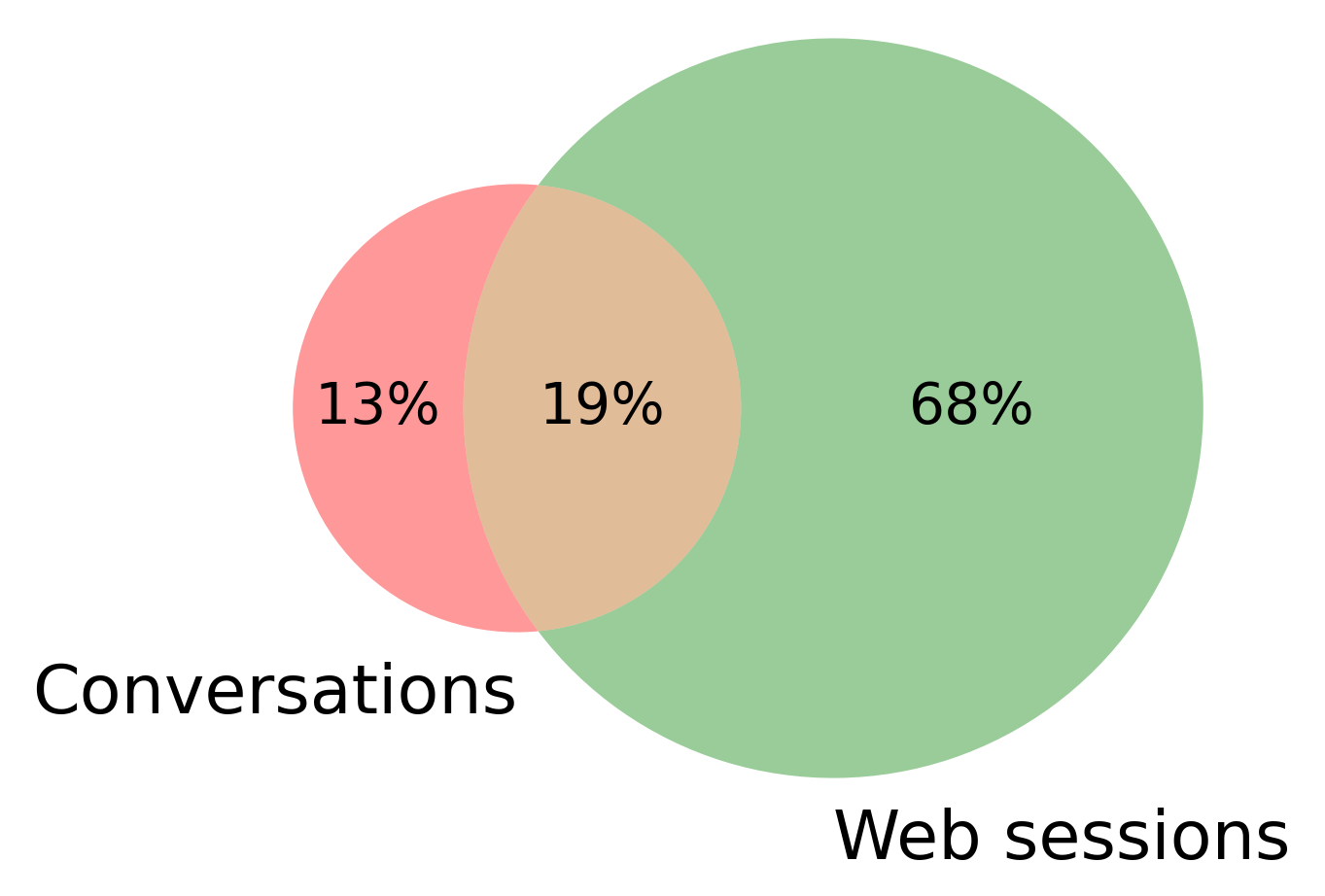}
    \caption{Distribution of the users having conversations and web sessions.}
    \label{fig:data_statistics}
\end{figure}

Table~\ref{tab:data_statistics} and Fig.~\ref{fig:data_statistics} show general statistics of the dataset after pre-processing. 
There are approximately $62$K purchases made by $52$K different users.
As observed from Fig.~\ref{fig:data_statistics}, not all users have had a conversation prior to their purchase (32\%). Likewise, not all users have had a web session prior to their purchase (87\%) why conversations and web sessions are naturally missing for part of the users. Only $19\%$ of the users have had both conversations and web sessions prior to their purchase.

\begin{figure*}[tb]
    \centering
    \includegraphics[width=0.6\textwidth]{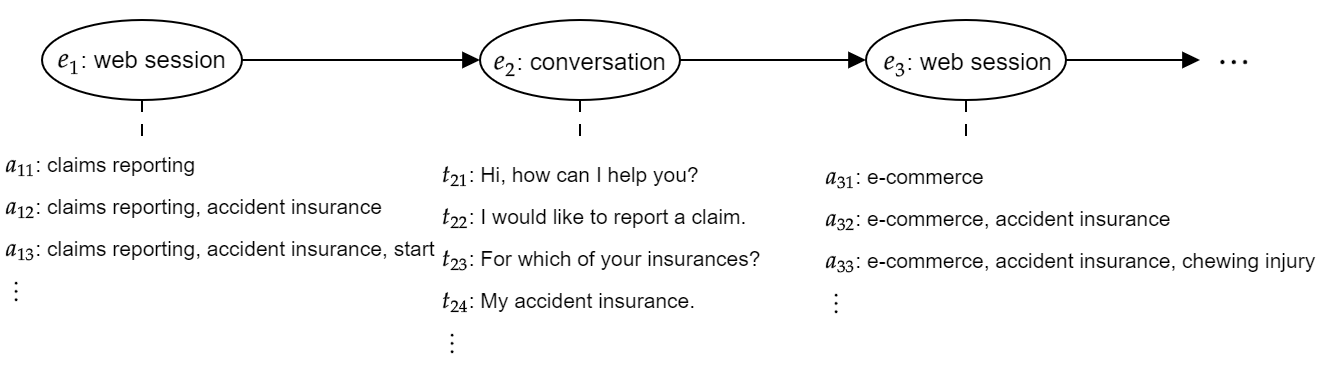}
    \vspace*{-\baselineskip}
    \caption{Example of a user's past events. An event can either be a \emph{conversation}, in the form of text, or a \emph{web session}, in the form of action tags.}
    \label{fig:user_history}
\end{figure*}

\section{Multi-modal Recommendations}

The core modeling issue 
consists of representing multi-modal user interactions. 
Next, we 
formalize the problem for this new task (Section~\ref{subsec:problem}),
present existing approaches (Section~\ref{subsec:existing}), and 
propose three models that specifically address the issue of naturally occurring incomplete modalities (Section~\ref{subsec:method}).
Together these models are meant to serve as a strong set of baselines for this new dataset.

\subsection{Problem Formalization}
\label{subsec:problem}
The goal of our RS is to recommend the next items that a user will buy, given the user's past conversations and web sessions. 
As opposed to existing multi-modal RSs, which deal with items with multi-modal representations, in our task: (1) we deal with multi-modal user interactions, and (2) parts of the modalities are naturally missing.
We collectively refer to web sessions and conversations as \emph{events} and represent a user as the list of the user's past events, $e_i$, chronologically ordered. Depending on the user's history, the event list can either exclusively contain conversations, web sessions, or a combination of both. In the latter case, the order of conversations and web sessions can vary for different users.
Moreover, the events have different modalities. A web session is a sequence of user actions, $\lbrace a_{i1},a_{i2},a_{i3},...,a_{in} \rbrace$, on the website, where an action is a set of tags. A conversation is a sequence of sentences, $\lbrace t_{i1},t_{i2},t_{i3},...,t_{in} \rbrace$, between a user and an insurance agent, in the form of text.
Fig.~\ref{fig:user_history} illustrates a user that has had a web session followed by a conversation, and finally a web session again.

The task is to learn a function, $f$, for predicting the probability that a user will buy each item $j$ after the last event $e_m$ based on the input sequence of a user's past events:
\begin{equation}
    f(e_1,e_2,e_3,...,e_m) = (\hat{p}_1,\hat{p}_2,\hat{p}_3,...,\hat{p}_J),
    \label{eq:task}
\end{equation}
where each element in $\lbrace e_1,e_2,e_3,...,e_m \rbrace$ can either be a conversation or web session, $\hat{p}_j$ is the estimated probability that item $j$ will be bought by the user, and $J$ is the total number of items.

\begin{figure*}[tb]
\centering
    \begin{tabular}{ccc}
        \includegraphics[width=0.2\textwidth]{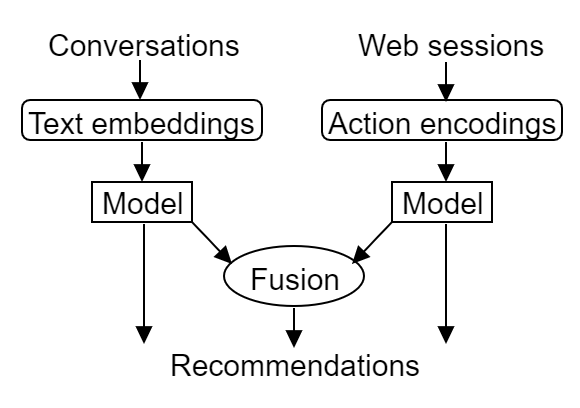} & 
        \includegraphics[width=0.2\textwidth]{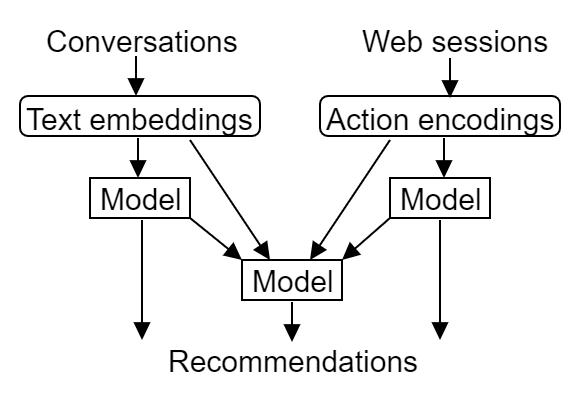} & 
        \includegraphics[width=0.2\textwidth]{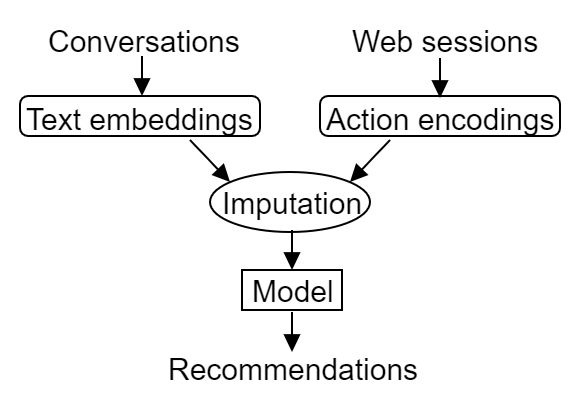} \\
        \textbf{(a) Late Fusion model} & \textbf{(b) Knowledge Distillation model} & \textbf{(c) Imputation models} \\
    \end{tabular}
    \vspace*{-0.5\baselineskip}
    \caption{Schematic overview of the baseline models.}
    \label{fig:process_late_fusion}
    \label{fig:process_knowledge_distillation}
    \label{fig:process_imputation}
\end{figure*}

\subsection{Existing Approaches}
\label{subsec:existing}
The following existing approaches can be applied to our problem.
\begin{itemize}[leftmargin=1em]
    \item \textbf{Popular} recommends the items with the largest number of purchases across users.
    \item \textbf{Conversation} is a model trained only on the conversations. For that, we use a neural text classifier that takes as input the average text embeddings of a user's past conversations and predicts the purchase probability of each item. Note that this model only provides recommendations for users that had conversations.
    \item \textbf{Web Session} is a model trained only on the web sessions, using the session-based RS presented in \citep{Bruun:2022:RecSys}, which takes into account the special characteristics of the insurance domain. Note that this model only provides recommendations for users that had web sessions.
    \item \textbf{Late Fusion} combines the output from the Conversation and Web Session models. It uses the recommendations from the Conversation model for those users that had only conversations and the recommendations from the Web Session model for those users that had only web sessions. Finally, it fuses the output from the two separate models for those users that have had both conversations and web sessions by averaging the predictions from the two models. This model is illustrated in Fig.~\ref{fig:process_late_fusion}a.
    \item \textbf{Knowledge Distillation} trains a joint model on users that had both conversations and web sessions. The model uses information from the two separate models as done in \citep{Wang:2020:KDD}. For users with only one modality, it uses the recommendations from the respective model for that modality; see Fig.~\ref{fig:process_knowledge_distillation}b.
    \item \textbf{Generative Imputation} trains a model that generates the missing modality from the other modality as done in \citep{Cai:2018:KDD, Tran:2017:CVPR, Wang:2018:EMNLP}. Once the missing modalities are imputed by the generative model, the modalities can be concatenated and jointly modeled. For a fair comparison, we use a neural network with the same architecture as the separate models. This model is illustrated in Fig.~\ref{fig:process_imputation}c.
    \item \textbf{Neutral Imputation} is similar to Generative Imputation. Because the modalities are not missing for technical reasons, we try a more neutral imputation strategy, so that fabricated conversations/web sessions do not corrupt the model. Missing conversations are imputed with the average text embedding in the training set and missing web sessions are imputed with the most frequent web session in the training set.
\end{itemize}

\subsection{Proposed Methods}
\label{subsec:method}
Next, we present novel models that aim to address the aspect of the problem concerning naturally induced incompleteness.

\subsubsection{Data Representation}
Our approach builds on top of the session-based RS proposed in \citet{Bruun:2022:RecSys} that takes into account the special characteristics of the insurance domain, namely (1) web sessions consist of various actions, not only interactions with items; (2) users can have multiple web sessions before the target action (i.e., purchase); and (3) the purchase occurs outside the web session.
We propose three different approaches to map the conversations into the same feature space as the web sessions and jointly model the two modalities with the session-based framework.
In the session-based model, a user is represented by the sequence of the user's past web sessions $\lbrace s_1,s_2,s_3,...,s_m \rbrace$.
A web session can be encoded in different ways. 
Based on~\citet{Bruun:2022:RecSys}, we encode a web session with a maximum pooling operation:
\begin{equation}
    s_i = \text{max}_{element}(a_{i1},a_{i2},a_{i3},...,a_{in}),
    \label{eq:encoding}
\end{equation}
where $\max_{element}(\cdot)$ is a function that takes the element-wise maximum of vectors, and $a_{ij}$ is a binarized vector of the action performed by a user.
Then, the ordered sequence of encoded sessions is passed through an RNN with gated recurrent units (GRU) that predicts what items the user will buy after the last time step.
We extend this framework to the multi-modal case, such that a user is now represented by the sequence of the user's past events $\lbrace e_1,e_2,e_3,...,e_m \rbrace$, where each event, $e_i$, can either be a conversation, $c_i$, or a web session, $s_i$. Next, we explain different ways of mapping the conversations and web sessions into the same feature space, so the events can be passed together through the RNN. In this way, a missing modality will not affect the model, since the events can be handled equally, once they are mapped into a common representation space.

\subsubsection{Keyword Model}
In the first way, called the \emph{Keyword} model, the conversations are represented by keywords extracted from the text. We then manually match the keywords with the actions from the web sessions. Each event is thereby a binarized sequence of keywords, that can be encoded in the same way as the actions in Eq.~\eqref{eq:encoding}, into a sequence of keyword vectors $\lbrace k_1,k_2,k_3,...,k_m \rbrace$ representing the events. Then, for every time step $i$ in the sequence of a user's events, an RNN with a single GRU layer computes the hidden state\footnote{Note that bias terms are omitted when hidden states are presented.}
\begin{align}
  \begin{aligned}
   h_i &= (1-z_i) \cdot h_{i-1} + z_i \cdot \hat{h}_i \\
   z_i &= \sigma(W_z k_i+U_z h_{i-1}), \\
   \hat{h}_i &= \tanh(W k_i+U(r_i \cdot h_{i-1})), \\
   r_i &= \sigma(W_r k_i + U_r h_{i-1}),
  \end{aligned}
  &&
  \begin{aligned}
   &\text{for}~~ i=1,..,m, \\ 
   &(update~~gate) \\
   &(candidate~~gate) \\
   &(reset~~gate)
  \end{aligned}
  \label{eq:GRU}
\end{align}
where $W_z, U_z, W, U, W_r$ and $U_r$ are weight matrices and $\sigma(\cdot)$ is the sigmoid function.
The Keyword model is illustrated in Fig.~\ref{fig:process}a, with the corresponding neural architecture shown in Figure~ \ref{fig:neural_architecture}a.

\subsubsection{Latent Feature Model}
\begin{figure}[tb]
\centering
    \begin{tabular}{cc}
        \includegraphics[width=0.4\columnwidth]{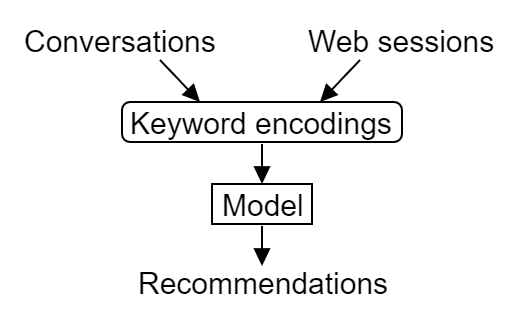} & \includegraphics[width=0.45\columnwidth]{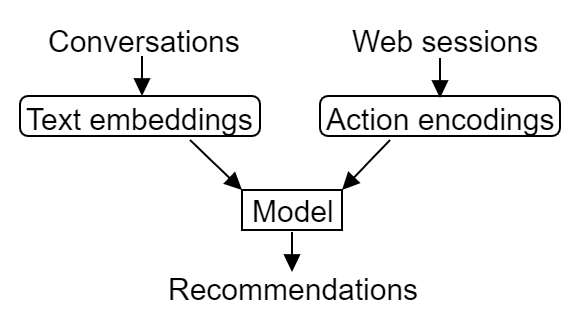} \\
        \textbf{(a) Keyword model} & \textbf{(b) Latent Feature model} \\
        \multicolumn{2}{c}{\includegraphics[width=0.65\columnwidth]{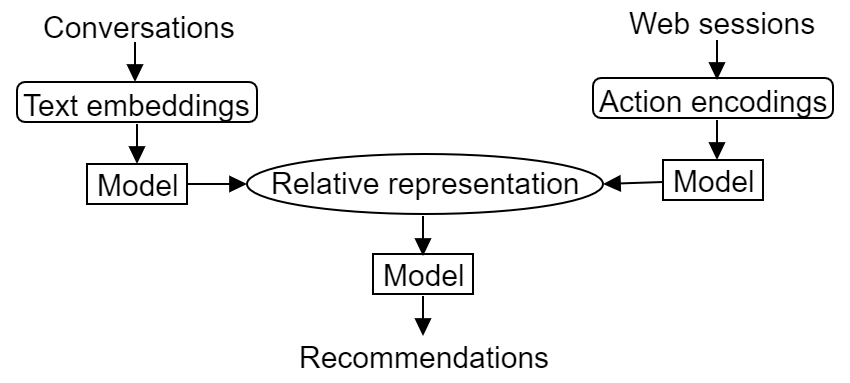}} \\
        \multicolumn{2}{c}{\textbf{(c) Relative Representation model}}
    \end{tabular}
    \vspace*{-0.5\baselineskip}
    \caption{Schematic overview of our models.}
    \label{fig:process}
    \vspace*{-0.5\baselineskip}
\end{figure}

\begin{figure}[tb]
\centering
    \begin{tabular}{cc}
        \includegraphics[width=0.35\columnwidth]{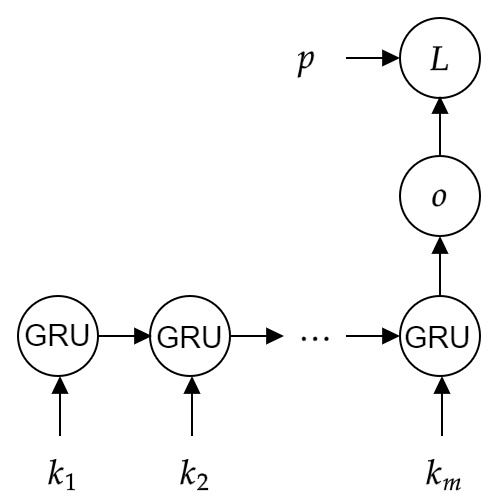} & \includegraphics[width=0.35\columnwidth]{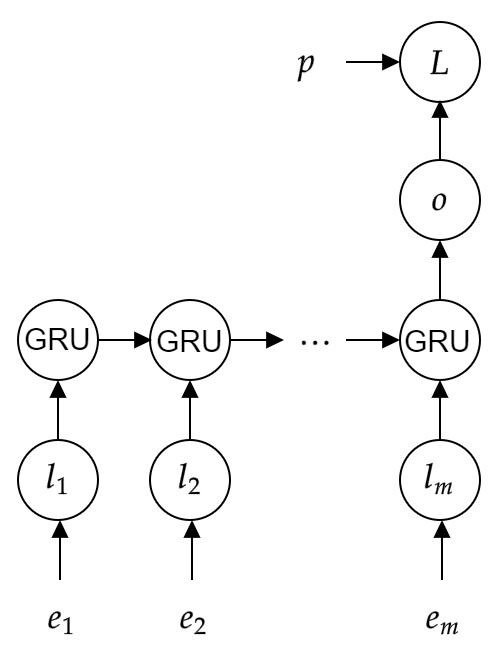} \\
        \textbf{(a) Keyword model} & \textbf{(b) Latent Feature model} \\
        \multicolumn{2}{c}{\includegraphics[width=0.65\columnwidth]{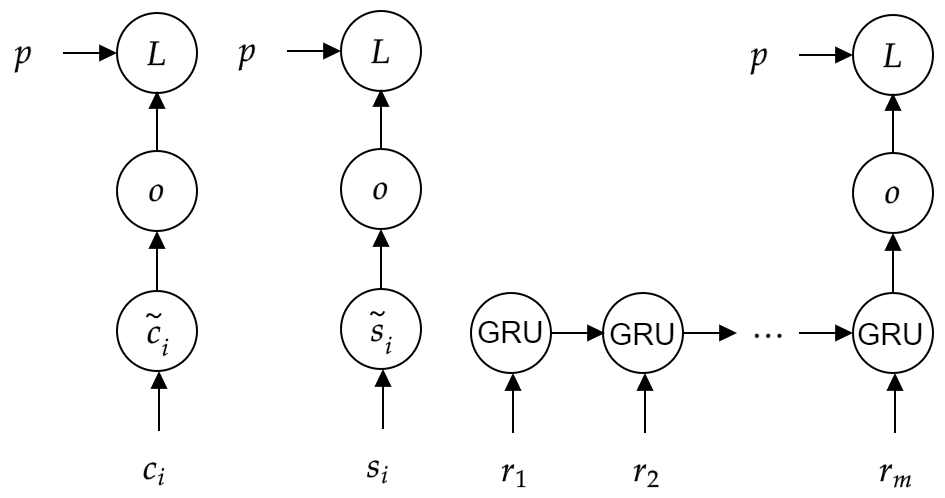}} \\
        \multicolumn{2}{c}{\textbf{(c) Relative Representation model}}
    \end{tabular}
    \vspace*{-0.5\baselineskip}
    \caption{Neural architectures of our models.}
    \label{fig:neural_architecture}
    \vspace*{-0.5\baselineskip}
\end{figure}

The second way of mapping the two modalities into a common feature space is called the \emph{Latent Feature} model. Here, conversations are represented by text embeddings, and web sessions are represented by action encodings (cf. Eq.~\eqref{eq:encoding}).
The two different representations are passed as input to the same neural network, where the first layer maps them into a common representation of latent features.
Formally, the following hidden state is computed for all the events in a user's sequence:
\begin{equation}
l_i = \tanh(W_c e_i \cdot \mathbbm{1}_{\{ e_i = c_i \}} + W_s e_i \cdot \mathbbm{1}_{\{ e_i = s_i \}}), ~\text{for}~i=1,..,m,
\label{eq:latent_feature}
\end{equation}
where $W_c$ and $W_s$ are weight matrices used to map the conversations and web sessions respectively into the vector, $l_i$, of common latent features, and $\mathbbm{1}_{\{\}}$ is an indicator function ensuring that $W_c$ is used whenever the event is a conversation and $W_s$ is used whenever the event is a session.
We use the hyperbolic tangent as the activation function for this hidden layer to promote the task of mapping the real-valued conversations and the binarized sessions into features on the same scale.
Then, for every time step in the sequence, $l_i$ is passed through the hidden state in Eq.~\eqref{eq:GRU} instead of $k_i$.
The weight matrices in Eq.~\eqref{eq:latent_feature} are optimized during the training of the full network. In that way, the latent features are automatically learned with respect to the task of generating recommendations.
The Latent Feature model is illustrated in Fig.~\ref{fig:process}b, with the neural architecture shown in Fig.~\ref{fig:neural_architecture}b.

In both cases, the neural network returns an output vector $o$ of length $J$ after the last time step. Because a user can buy multiple items at the same time, the learning task is considered a multi-label classification, and the sigmoid function is used on each element of $o$ as the output activation function to compute the likelihood of purchase:
\begin{equation}
    \hat{p}_j = \sigma(o_j), ~~\text{for}~~ j=1,\ldots,J.
    \label{eq:output}
\end{equation}
During training, the loss function is computed by comparing $\hat{p}$ with the binarized vector of the items purchased, $p$. Due to the learning task being multi-label classification, the loss function is defined as the sum of the binary cross-entropy loss over all items:
\begin{equation}
    L = - \sum_{j=1}^J \Big( p_j \cdot \log (\hat{p}_j) +(1-p_j) \cdot \log (1-\hat{p}_j) \Big).
    \label{eq:loss_function}
\end{equation}

\subsubsection{Relative Representation Model}
We propose a third way, called the \emph{Relative Representation} model. We use 
relative representation~\citep{Moschella:2023:ICLR} 
that encodes the intrinsic information of a dataset learned by a neural network. Each data point becomes a set of coefficients that encode the point as a function of other data samples. It thereby enables the comparison of latent spaces across different neural networks.
Hence, we train two separate neural networks from where latent representations of the two modalities can be extracted. We then use the method in \cite{Moschella:2023:ICLR} to make the two representations relative and thereby comparable to each other. The conversations and web sessions can now jointly be modeled by using their relative representations.
Particularly, the two separate networks take individual conversations/web sessions as input (as opposed to the users' sequences of conversations/web sessions) and have an intermediate hidden state that can be used to create latent representations of single conversations/web sessions. They are trained on the same downstream recommendation task.
Formally, a conversation, $c_i$, is represented by text embeddings and passed through a dense layer that computes a latent representation
\begin{equation}
    \tilde{c}_i = \tanh(W_c c_i) \label{eq:latent_conv},
\end{equation}
where $W_c$ is a weight matrix. As in the Latent Feature model, we use the hyperbolic tangent as the activation function to promote the transformation of the different modalities into a common latent space.
This representation is then used to solve the downstream recommendation task by passing it through another dense layer that computes the output $o$ to be used in Eq.~\eqref{eq:output}:
\begin{equation}
    o = W_o \tilde{c}_i,
\end{equation}
where $W_o$ is another weight matrix.
A web session, $s_i$, is represented by action encodings and similarly transformed into a latent representation, $\tilde{s}_i$, that can be used to solve the downstream recommendation task:
\begin{align}
    &\tilde{s}_i = \tanh(W_s s_i), \label{eq:latent_ses} \\
    &o = W_o \tilde{s}_i.
\end{align}
Once the weights are optimized with respect to the recommendation task, the first part of the networks (i.e., Eq.~\eqref{eq:latent_conv} and \eqref{eq:latent_ses}) are used to convert conversations and web sessions into latent representations.
Following the method in \cite{Moschella:2023:ICLR}, a subset, $\mathcal{X}$, of the training set, denoted anchors, is selected. In our case, the anchors are selected among the users that are represented by both modalities. Given the indices of the anchor users in an arbitrary ordering $x_1,x_2,\ldots,x_{|\mathcal{X}|}$, the relative representation, $r_i$, of a conversation is computed as a function of the anchor users' conversations:
\begin{equation}
    r_i = (sim(\tilde{c}_i, \tilde{c}_{x_1}),sim(\tilde{c}_i, \tilde{c}_{x_2}),\ldots,sim(\tilde{c}_i, \tilde{c}_{x_{|\mathcal{X}|}})),
    \label{eq:relative_representation_conversation}
\end{equation}
where $sim()$ is a similarity function yielding a scalar score.
Similarly, the relative representation of a session is given by
\begin{equation}
        r_i = (sim(\tilde{s}_i, \tilde{s}_{x_1}),sim(\tilde{s}_i, \tilde{s}_{x_2}),\ldots,sim(\tilde{s}_i, \tilde{s}_{x_{|\mathcal{X}|}})).
        \label{eq:relative_representation_session}
\end{equation}
A sequence of events is now represented by $\lbrace r_1,r_2,\ldots,r_m \rbrace$, where each $r_i$ is either the relative representation of a conversation or a web session. The two modalities are jointly modeled by passing $r_i$ through the hidden state in Eq.~\eqref{eq:GRU} instead of $k_i$ for every time step in the sequence. This model is shown in Figs.~\ref{fig:process}c and \ref{fig:neural_architecture}c.



\section{Experimental Setup}

\label{sec:expsetup}
First, we describe the evaluation procedure, then the baselines, implementation details, and hyperparameter tuning.

\subsection{Evaluation Procedure}
As a test set, we use the latest $10\%$ of purchases with associated past conversations and web sessions. The remaining $90\%$ is used for training.

The models generate a score for how likely the user will buy each item, which is then sorted as a ranked list. There are two types of items: new insurance products and additional coverage.
Since it is only possible for a user to buy additional coverage if the user has the corresponding base insurance product, we use a post filter to set the score to the lowest score if that is not the case, as per~\citet{Aggarwal:2016:book}.
The list of ranked items is evaluated with Hit Rate (HR) and Mean Average Precision (MAP).
Besides reporting HR and MAP averaged across all users in the test set (the union), we further break down the performance by the users that have only had conversations (the conversations-only), the users that have only had web sessions (the web sessions-only), and the users that have both had conversations and web sessions (the intersection), since this is of interest to our research questions. Because some of these subsets of users are relatively small (e.g., only 13\% conversations-only), we report the average of the performance measures over five models trained from different seeds to account for randomness.
We use a cutoff threshold of three because (1) the total number of items is $24$, therefore high cut-offs (e.g., $\geq10$) will not inform on the actual quality of the RSs; (2) on the user interface the user will be recommended up to three items.
Additionally, we report MAP scores for all cut-off values from one to five.
Experimental results are supported by statistical testing. For HR we use McNemar's test \citep{Dietterich:1998:NC} and for all other measures we use one-way ANOVA \citep{Kutner:2005:book}, both with a confidence level of $0.05$.

\subsection{Implementation \& Hyperparameters}

\begin{table}[tb]
\centering
\caption{Hyperparameters}
\label{tab:hyperparameters}
\vspace*{-0.5\baselineskip}
\resizebox{0.65\columnwidth}{!}{%
\begin{tabular}{lrrr}
\hline
\multicolumn{1}{c}{Model} & \multicolumn{1}{c}{Batch size} & \multicolumn{1}{c}{Units} & \multicolumn{1}{c}{Dropout} \\ \hline
Conversation & 512 & 64 & 0.2 \\
Web Session & 256 & 256 & 0.3 \\
Knowledge Distillation & 256 & 128 & 0.4 \\
Generative Imputation & 128 & 256 & 0.2 \\
Neutral Imputation & 128 & 128 & 0.2 \\
Keyword & 512 & 256 & 0.2 \\
Latent Feature & 512 & 256 & 0.3 \\
Relative Representation & 256 & 256 & 0.3 \\ \hline
\end{tabular}%
}
\end{table}

All implementation is in \texttt{Python} \texttt{3.9.13} and \texttt{TensorFlow} \texttt{2.10.0}. We used \texttt{Adam} as the optimizer with TensorFlow's default settings for the learning rate, exponential decay rates, and the epsilon parameter. Early stopping was used to choose the number of epochs based on the minimum loss on the validation dataset. We used two-layer networks\footnote{In all models the second layer is a dense layer with ReLU activation function.} with dropout regularization on the first hidden layer.
We partitioned the training set in the same way as the whole dataset, so the validation set includes the latest $10\%$ of purchases with associated conversations/web sessions, and the remaining is used for training.
Depending on the model, we remove action tags or keywords that have a frequency of less than $0.1$ percent.
We tuned the hyperparameters of each neural model (batch size, number of units, and dropout rate) on the validation set using grid search. We test powers of two for the batch size and number of units ranging from 64 to 256. For the dropout rate, we test values in $\lbrace 0.2, 0.3, 0.4 \rbrace$.
The final hyperparameters used are reported in Table~\ref{tab:hyperparameters}.
In the Knowledge Distillation model, we use as distillation loss the sum of the binary cross-entropy loss over all items, as we are dealing with a multi-label classification task. The $\alpha$ and $\beta$ parameters, which are used to control how much knowledge the student model gets from the teacher models, are set in proportion to how much data the student models are trained on. That is $\alpha=0.32$ for the Conversation model and $\beta=0.87$ for the Web Session model. In the Relative Representation model, we use cosine similarity as the similarity measure (see Eqs.~\eqref{eq:relative_representation_conversation}~and~\eqref{eq:relative_representation_session}). We tune the number of anchors on the validation set and find the optimal number to be 125 which are sampled from each class in the training set.
\section{Results}

\begin{table*}[tb]
\centering
\caption{Performance results. All results marked with * are significantly different from the Latent Feature model.
The best scores for each measure are boldfaced. Percentages in brackets denote relative differences w.r.t. the strongest baseline (Imputation).}
\vspace*{-0.5\baselineskip}
\label{tab:results1}
\resizebox{0.9\textwidth}{!}{%
\begin{tabular}{lrrrrrrrr}
\hline
\multicolumn{1}{c}{\multirow{2}{*}{Model}} & \multicolumn{2}{c}{Union} & \multicolumn{2}{c}{Conversations-only} & \multicolumn{2}{c}{Web Sessions-only} & \multicolumn{2}{c}{Intersection} \\ \cline{2-9} 
\multicolumn{1}{c}{} & \multicolumn{1}{c}{HR@3} & \multicolumn{1}{c}{MAP@3} & \multicolumn{1}{c}{HR@3} & \multicolumn{1}{c}{MAP@3} & \multicolumn{1}{c}{HR@3} & \multicolumn{1}{c}{MAP@3} & \multicolumn{1}{c}{HR@3} & \multicolumn{1}{c}{MAP@3} \\ \hline
Popular & 0.4595* & 0.2742* & 0.5249* & 0.3318* & 0.4287* & 0.2424* & 0.5111* & 0.3340* \\
Conversation & - & - & 0.6623* & 0.4975 & - & - & 0.6184* & 0.4497* \\
Web Session & - & - & - & - & 0.6642* & 0.5188 & 0.6512* & 0.4750* \\
Late Fusion & 0.6703* & 0.5179* & 0.6623* & 0.4975 & 0.6642* & 0.5188 & 0.6949* & 0.5287 \\
Knowledge Distillation & 0.6697* & 0.5168* & 0.6623* & 0.4975 & 0.6642* & 0.5188 & 0.6921* & 0.5235* \\
Generative Imputation & 0.6685* & 0.5156* & 0.6582* & 0.4878* & 0.6635* & 0.5180 & 0.6910* & 0.5270 \\
Neutral Imputation & 0.6767* & 0.5161* & 0.6843* & 0.5053 & 0.6687 & 0.5151 & 0.6966* & 0.5265* \\ \hline
Keyword & 0.6840 (1.08\%) & 0.5250 (1.74\%) & 0.6940 (1.42\%) & 0.5198 (2.86\%) & 0.6681* (-0.09\%) & 0.5141 (-0.18\%) & \textbf{0.7270 (4.37\%)} & \textbf{0.5629 (6.91\%)} \\
Latent Feature & \textbf{0.6872 (1.55\%)} & \textbf{0.5339 (3.46\%)} & \textbf{0.7145 (4.42\%)} & \textbf{0.5415 (7.17\%)} & \textbf{0.6712 (0.38\%)} & \textbf{0.5238 (1.69\%)} & 0.7183 (3.12\%) & 0.5603 (6.44\%) \\
Relative Representation & 0.6846 (1.16\%) & 0.5308 (2.85\%) & 0.7105 (3.84\%) & 0.5369 (6.25\%) & 0.6696 (0.14\%) & 0.5216 (1.27\%) & 0.7138 (2.46\%) & 0.5554 (5.49\%) \\ \hline
\end{tabular}%
}
\end{table*}

\begin{figure*}[tb]
    \centering
    \begin{subfigure}{0.24\textwidth}
        \centering
        \includegraphics[width=\columnwidth]{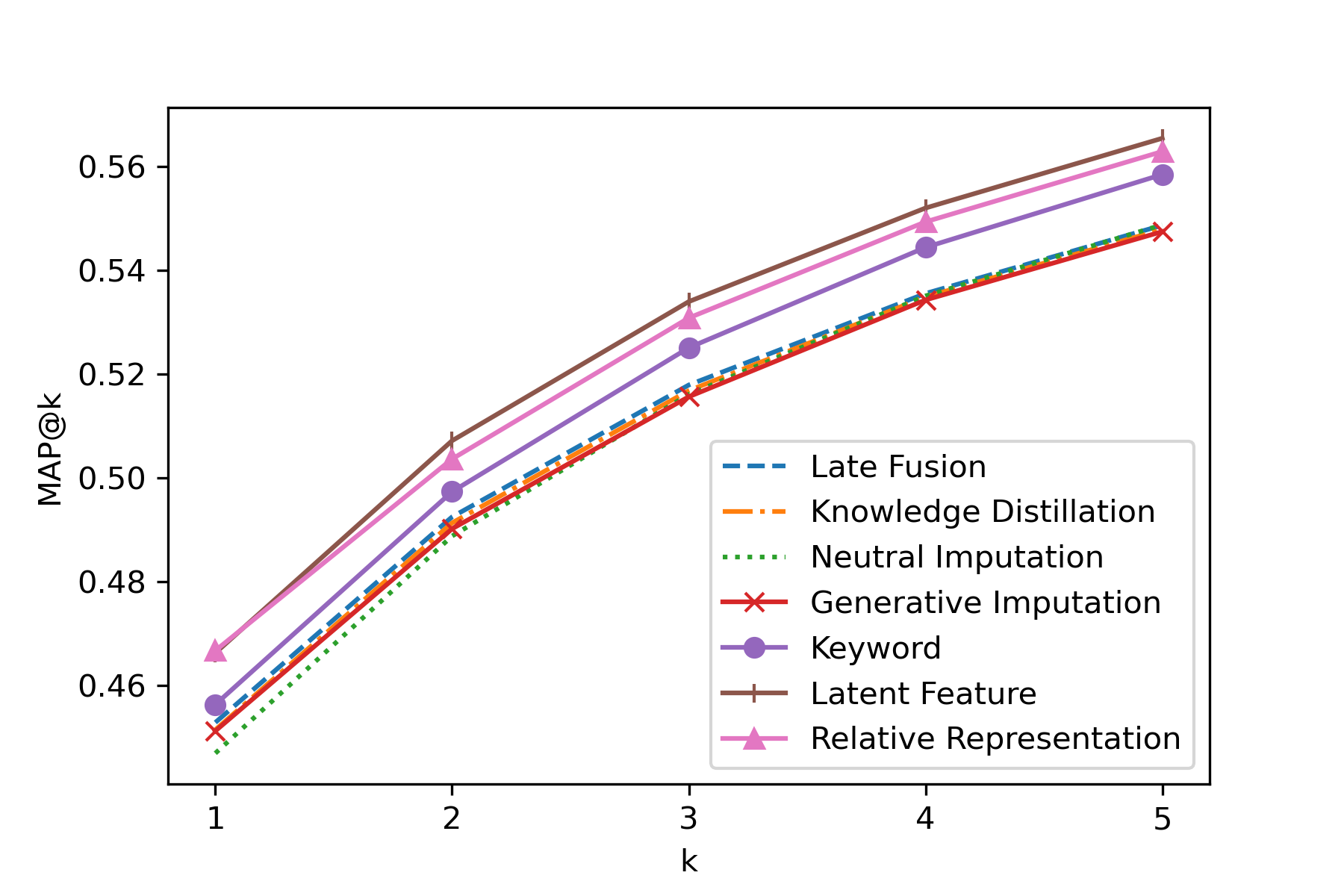}
        \caption{Union}
    \end{subfigure}
    \begin{subfigure}{0.24\textwidth}
        \centering
        \includegraphics[width=\columnwidth]{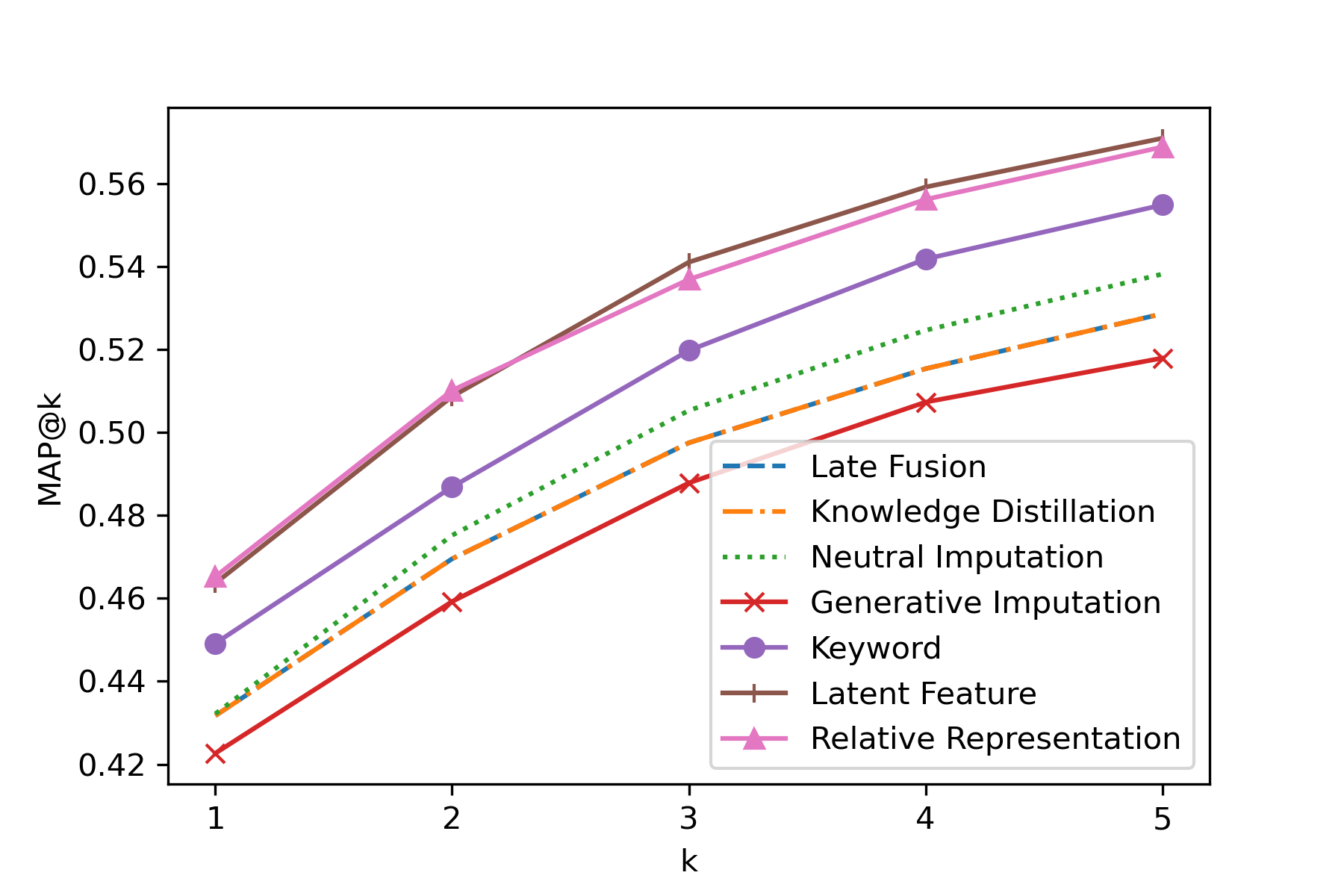}
        \caption{Conversations-only}
    \end{subfigure}
    \begin{subfigure}{0.24\textwidth}
        \centering
        \includegraphics[width=\columnwidth]{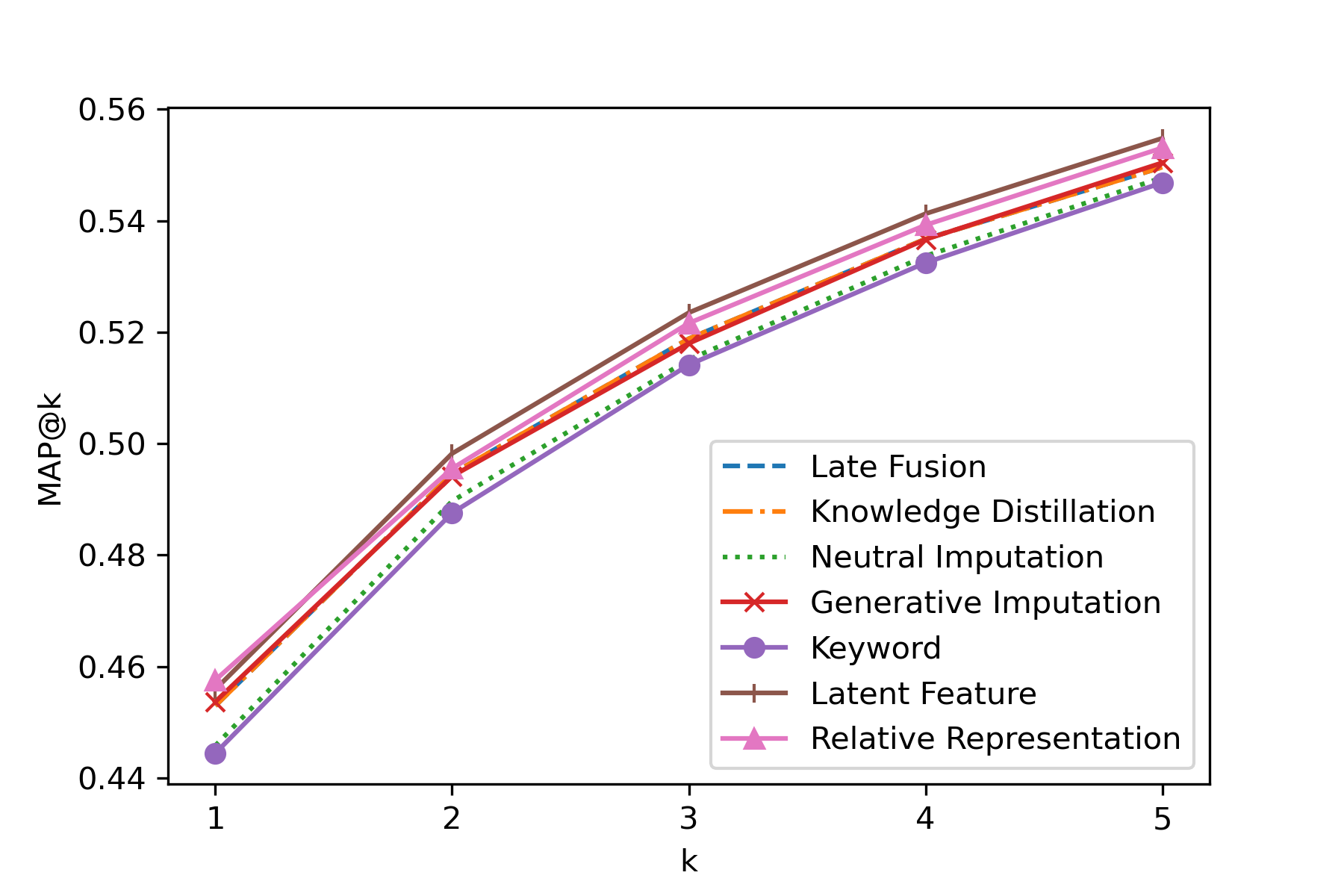}
        \caption{Web Sessions-only}
    \end{subfigure}
    \begin{subfigure}{0.24\textwidth}
        \centering
        \includegraphics[width=\columnwidth]{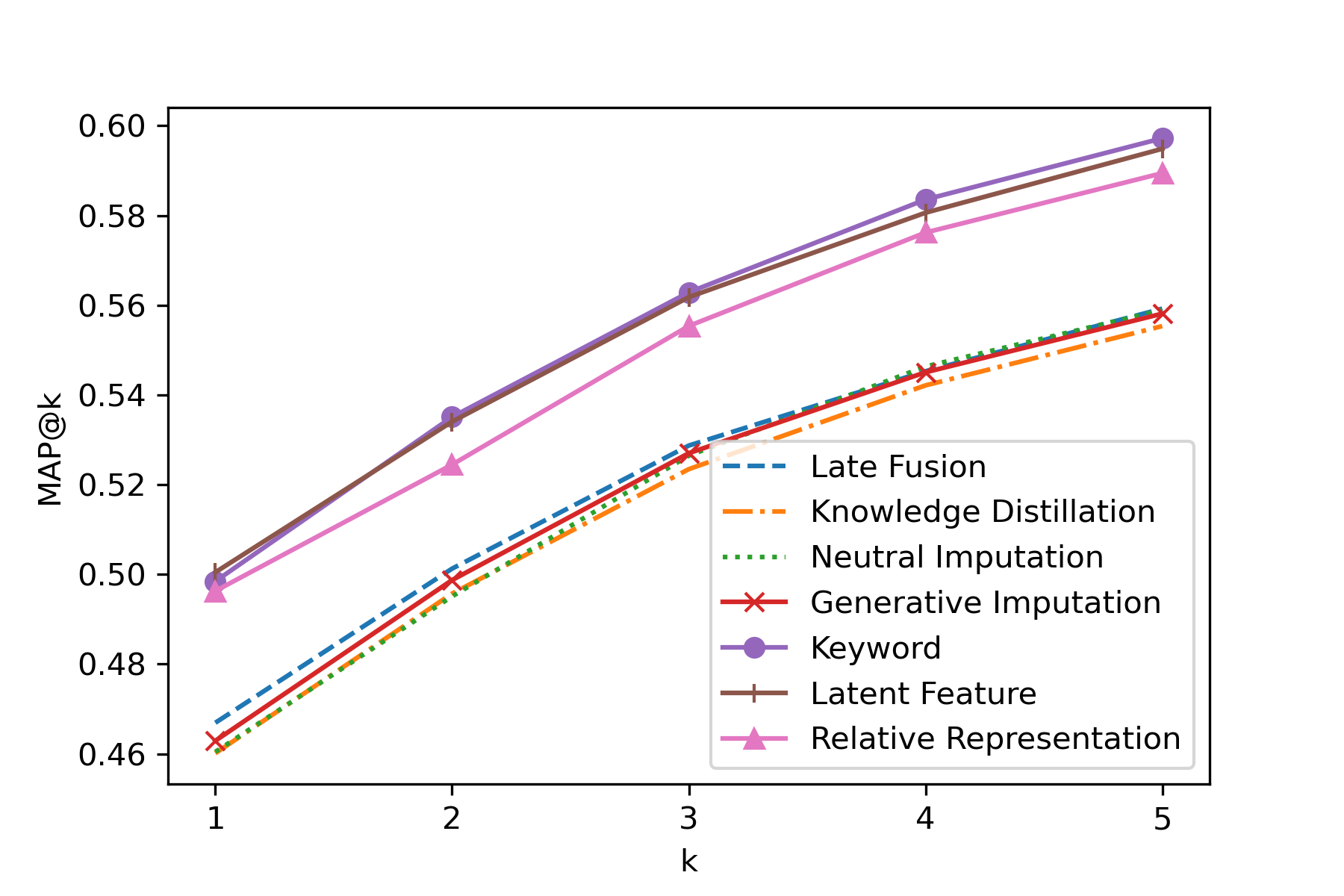}
        \caption{Intersection}
    \end{subfigure}
    \vspace*{-0.5\baselineskip}
    \caption{MAP@k for varying choices of the cutoff threshold $k$.}
    \label{fig:varying_thresholds}
\end{figure*}

\label{sec:results}
Next, we compare the different approaches presented in Section~\ref{subsec:method} in order to answer our research questions.
Table~\ref{tab:results1} presents performance results.
Figure ~\ref{fig:varying_thresholds} shows MAP at varying cutoffs $k$.  
We have similar results for HR which are omitted to save space.

\begin{itemize}
    \item[\textbf{RQ1}] \textbf{How can we effectively learn recommendations from multi-modal user interactions?} 
\end{itemize}
We look at the overall results of the models on the union (i.e., the entire dataset).
The results show that even though there are few items, all the models outperform the simple Popular baseline considerably, showing there is information in the multiple modalities that can be learned.

Simply averaging the predictions from the two separate models, as is done in the Late Fusion model, proves to be a strong baseline; we observe that the Knowledge Distillation and Imputation models do not manage to improve over the Late Fusion model by jointly modeling the two modalities.
It is reasonable that the training data is too small for the Knowledge Distillation model which is trained on the intersection data alone, even though it uses information from the two separate models.
For the Imputation models, especially the Generative Imputation model, it is likely that the synthetic imputation adds too much noise to the data.

The results show that our approach of mapping the two modalities into the same feature space significantly outperforms all the existing methods with the Latent Feature model being the best.

\begin{itemize}
    \item[\textbf{RQ2}] \textbf{How does it affect the quality of recommendations to jointly model the multiple modalities compared to separately?} 
\end{itemize}
We look further into the results broken down by conversations-only, web sessions-only, and intersection.
We observe that the Web Session model performs better than the Conversation model on the intersection. This is likely because there are considerably more web sessions than conversations in the training data.
Note that the Late Fusion and the Knowledge Distillation models use the recommendations from the separate models on the conversations-only and web sessions-only, hence the identical performance in these cases.

We find that the strength of the Late Fusion model is due to a considerable improvement on the intersection compared to the Conversation and Web Session models. This shows that the modalities capture different aspects of the problem.
The results further show that the Keyword, Latent Feature and Relative Representation models successfully manage to model important interactions between the two modalities, while this is not the case for the Knowledge Distillation and Imputation models, as performance increases considerably on the intersection compared to the baselines, with the Keyword model being best.

The Neutral Imputation model improves slightly on the con\-ver\-sa\-tions-only subset, suggesting that the smaller conversation data benefits from the larger web session data when jointly modeling the two modalities. This does not apply to the Generative Imputation model, even though the modalities are also jointly modeled with this model.
We observe that the Keyword, Latent Feature and Relative Representation models all improve considerably on the conversations-only subset while preserving the same performance on the web sessions-only subset compared to the existing methods. It shows that the small conversation dataset successfully benefits from the larger web session data when mapping the two modalities into the same feature space. The improvement is greatest with the Latent Feature and Relative Representation models. It is likely due to a loss of information when representing the conversations by keywords compared to text embeddings.

Figure~\ref{fig:varying_thresholds} shows that the results are consistent across various cutoff thresholds. Across all cutoff values, there is a clear gap between our proposed models and the existing methods with the exception of the Keyword model on the web sessions-only where performance does not improve over the existing methods.

Note that our models do not have more steps, more parameters or longer training times than the existing models, why they are not more computationally expensive.
\section{Analysis}
\label{sec:analysis}
We conduct further analysis to understand what makes this dataset and problem challenging.
We break down the performance of our proposed models to understand the impact of the number of past conversations and web sessions as well as the order of them. Then we use t-SNE to visualize the representations of the modalities before and after they are passed through the neural models.

\subsection{Number of Events}

\begin{figure*}[tb]
    \vspace*{-0.5\baselineskip}
    \centering
    \begin{subfigure}{0.24\textwidth}
        \centering
        \includegraphics[width=\columnwidth]{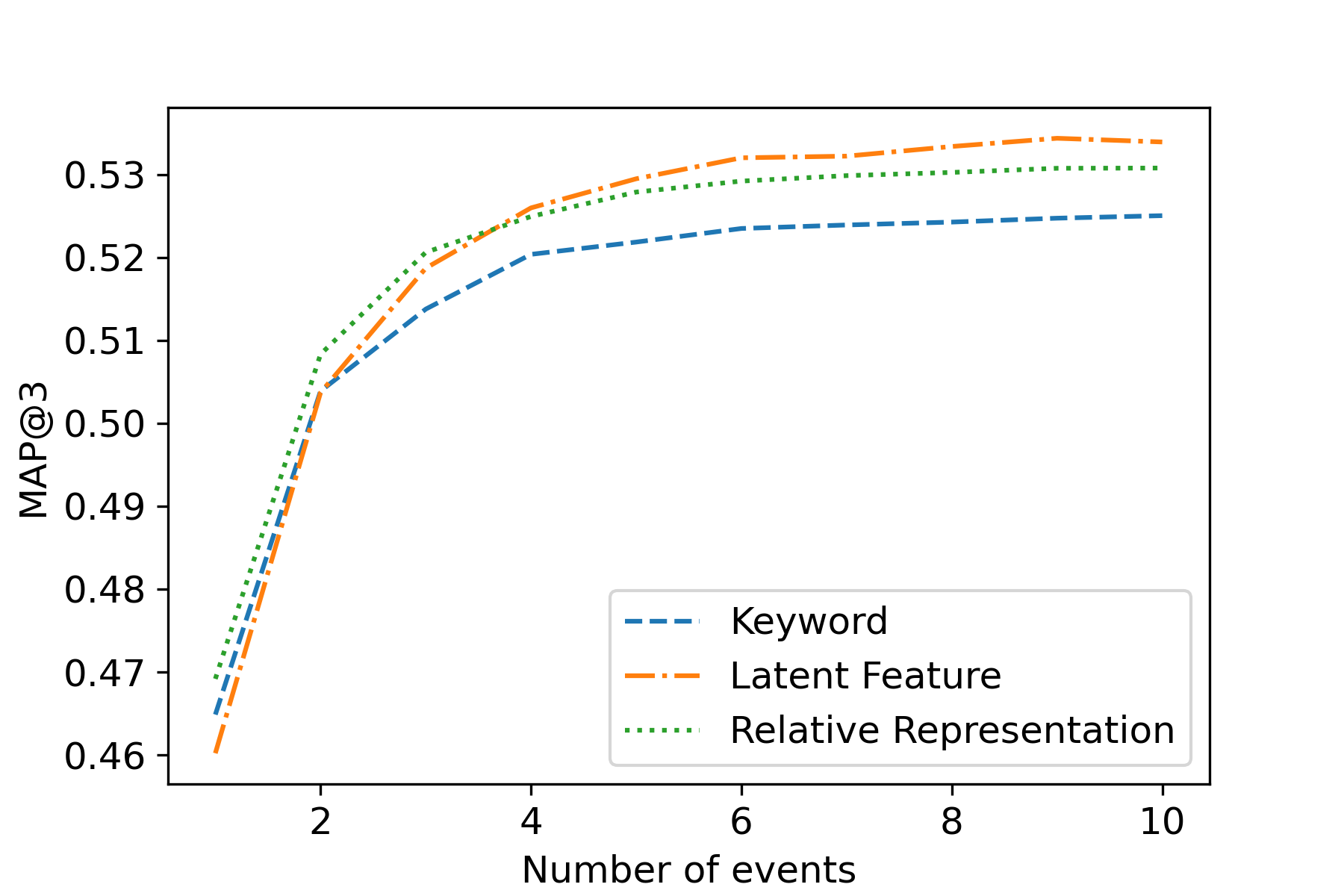}
        \caption{Union}
    \end{subfigure}
    \begin{subfigure}{0.24\textwidth}
        \centering
        \includegraphics[width=\columnwidth]{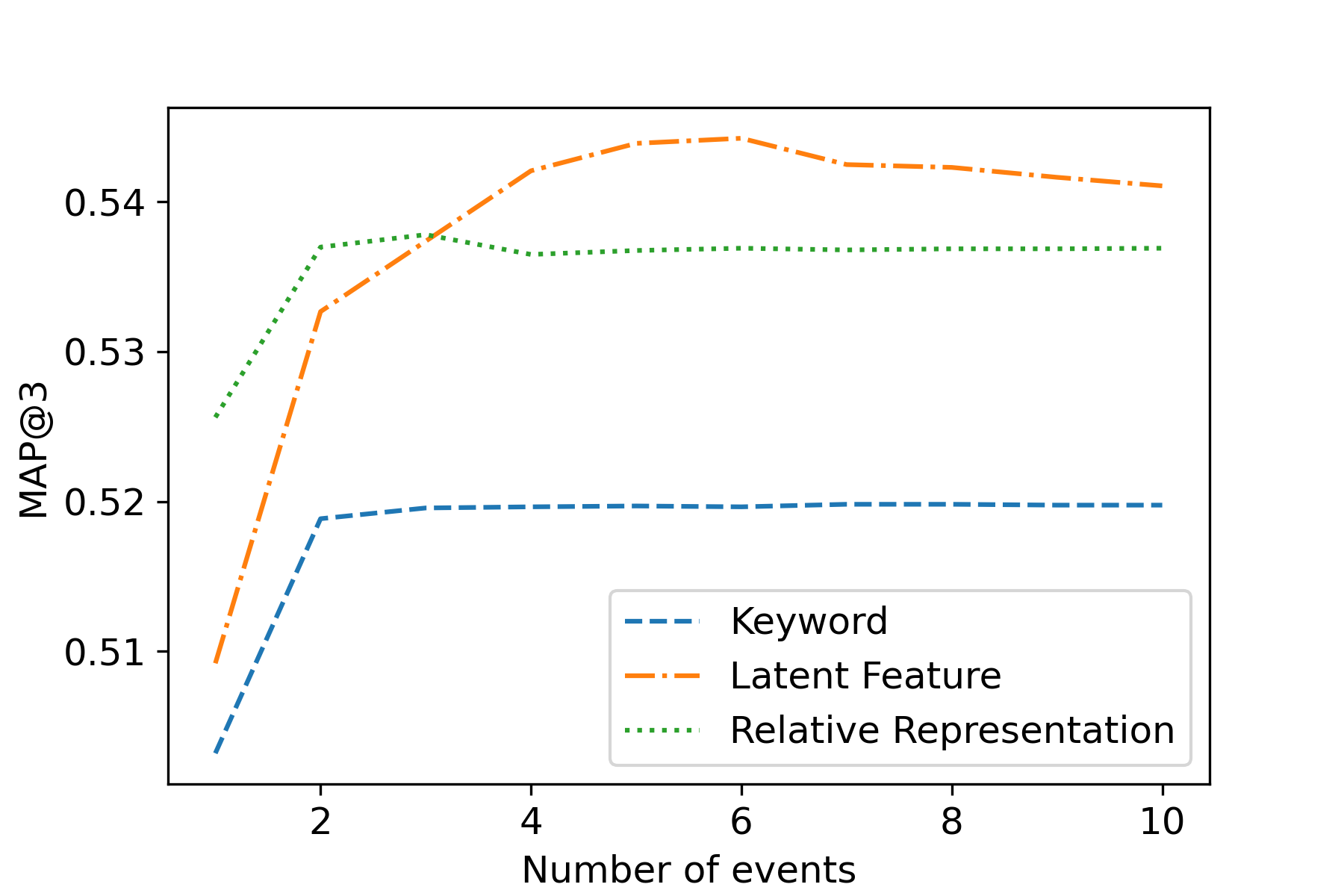}
        \caption{Conversations-only}
    \end{subfigure}
    \begin{subfigure}{0.24\textwidth}
        \centering
        \includegraphics[width=\columnwidth]{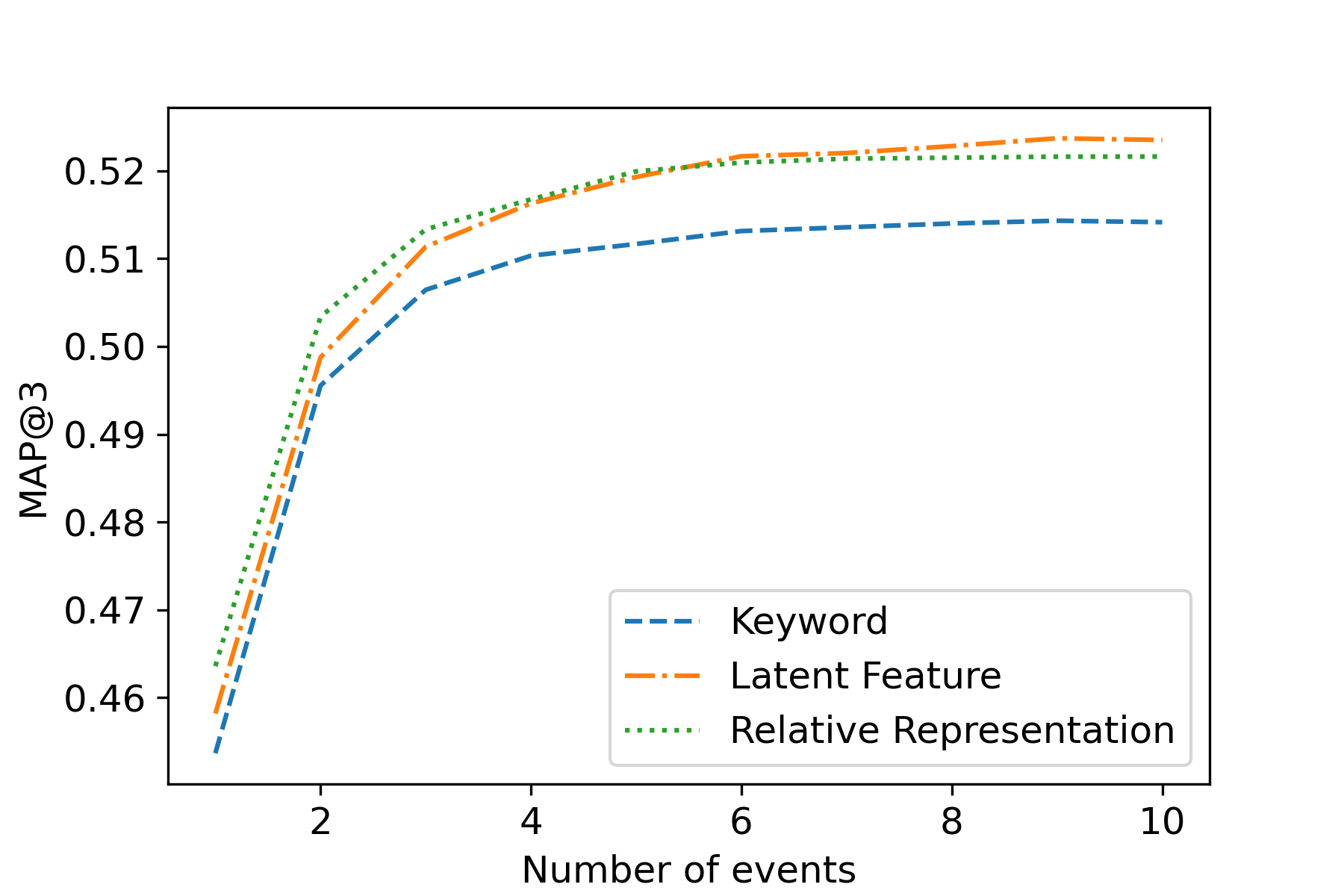}
        \caption{Web Sessions-only}
    \end{subfigure}
    \begin{subfigure}{0.24\textwidth}
        \centering
        \includegraphics[width=\columnwidth]{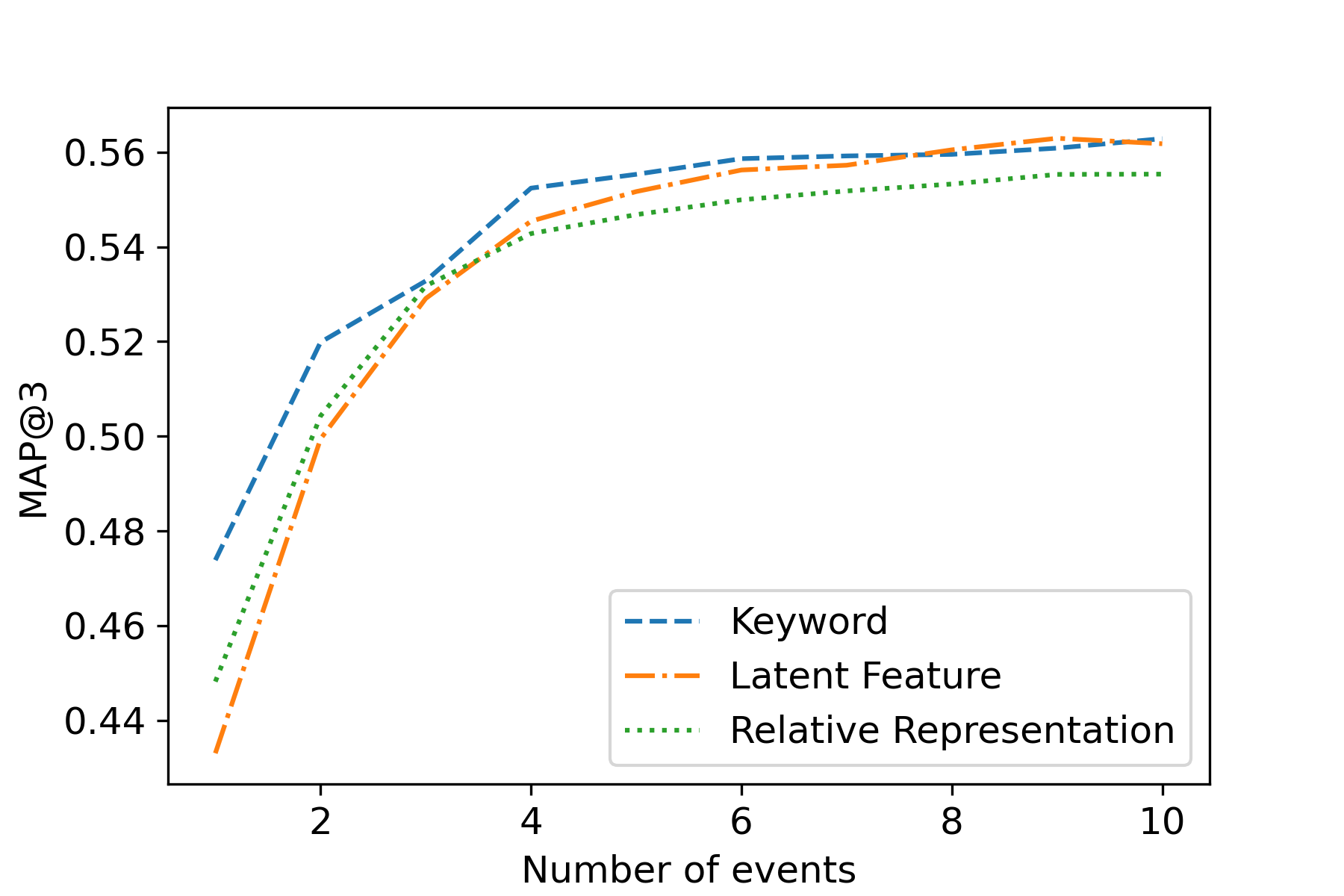}
        \caption{Intersection}
    \end{subfigure}
    \vspace*{-0.5\baselineskip}
    \caption{MAP@3 for a different number of events.}
    \label{fig:n_events}
\end{figure*}

We analyze how the number of events (i.e., conversations and web sessions) affects our models.
Figure~\ref{fig:n_events} shows MAP@3 broken down by the number of events, starting with only the most recent event, up to including all the available events.
In general, performance increases with the number of events up to about five events, after which it flattens out.
It is less important for users with only conversations to include many historical events. Particularly, the Keyword and Relative Representation models do not benefit from more than two/three conversations.
The Relative Representation model generally outperforms the two others when only a few events are included, except for the intersection where the Keyword model is best.
It is likely because the relative representations are
computed individually for each conversation and web session, making it less dependent on the whole user history.
We observe similar results for HR, which are not included due to space constraints.

\subsection{Event Order}

\begin{table*}[tb]
\centering
\caption{Study of the event order. Relative changes are in parentheses.}
\label{tab:event_order}
\vspace*{-0.5\baselineskip}
\resizebox{0.9\textwidth}{!}{%
\begin{tabular}{llrrrrrrrr}
\hline
\multicolumn{2}{c}{\multirow{2}{*}{Model}} & \multicolumn{2}{c}{Union} & \multicolumn{2}{c}{Conversations-only} & \multicolumn{2}{c}{Sessions-only} & \multicolumn{2}{c}{Intersection} \\ \cline{3-10} 
\multicolumn{2}{c}{} & \multicolumn{1}{c}{HR@3} & \multicolumn{1}{c}{MAP@3} & \multicolumn{1}{c}{HR@3} & \multicolumn{1}{c}{MAP@3} & \multicolumn{1}{c}{HR@3} & \multicolumn{1}{c}{MAP@3} & \multicolumn{1}{c}{HR@3} & \multicolumn{1}{c}{MAP@3} \\ \hline
Keyword & original event order & 0.6840 & 0.5250 & 0.6940 & 0.5198 & 0.6681 & 0.5141 & 0.7270 & 0.5629 \\
 & shuffled event order & 0.6775 (-0.95\%) & 0.5216 (-0.65\%) & 0.6935 (-0.07\%) & 0.5181 (-0.34\%) & 0.6645 (-0.54\%) & 0.5154 (0.26\%) & 0.7073 (-2.71\%) & 0.5433 (-3.49\%) \\ \hline
Latent Feature & original event order & 0.6872 & 0.5339 & 0.7145 & 0.5415 & 0.6712 & 0.5238 & 0.7183 & 0.5603 \\
 & shuffled event order & 0.6722 (-2.18\%) & 0.5164 (-3.27\%) & 0.7135 (-0.14\%) & 0.5300 (-2.13\%) & 0.6610 (-1.52\%) & 0.5133 (-2.01\%) & 0.6791 (-5.45\%) & 0.5171 (-7.71\%) \\ \hline
Relative Representation & original event order & 0.6846 & 0.5308 & 0.7105 & 0.5369 & 0.6696 & 0.5216 & 0.7138 & 0.5554 \\
 & shuffled event order & 0.6776 (-1.03\%) & 0.5232 (-1.43\%) & 0.6992 (-1.59\%) & 0.5314 (-1.03\%) & 0.6664 (-0.49\%) & 0.5183 (-0.63\%) & 0.6978 (-2.23\%) & 0.5329 (-4.05\%) \\ \hline
\end{tabular}%
}
\end{table*}

We analyze the importance of event order by randomly shuffling the order of events and retraining the models. We shuffle the order in both training and test data and perform the experiment five times to account for randomness. The mean performance is presented in Table~\ref{tab:event_order}.
We observe that the performance of the Keyword model drops less than that of the Latent Feature and Relative Representation models on the union (less than -0.95\%). It shows that the Latent Feature and Relative Representation models are more successful in capturing dependencies in the sequential order of events. In addition, it is also likely that these models overfit the data, as they have more parameters than the Keyword model.
The Latent Feature model is more affected than the Relative Representation model when shuffling the order. It is likely because the relative representations are computed individually for each conversation and web session while the latent features are learned from the users' sequences of conversations/web sessions.
For all models, the biggest drop in performance is seen on the intersection (up to $-7.71\%$). It explains part of the superiority of mapping the modalities into a common feature space since sequential dependencies across the two modalities prove to be important.



\subsection{Visualization}

\captionsetup[subfigure]{aboveskip=1pt} 
\begin{figure*}[tb]
    \vspace*{-0.5\baselineskip}
    \centering
    \begin{subfigure}{0.25\textwidth}
        \centering
        \includegraphics[width=\columnwidth]{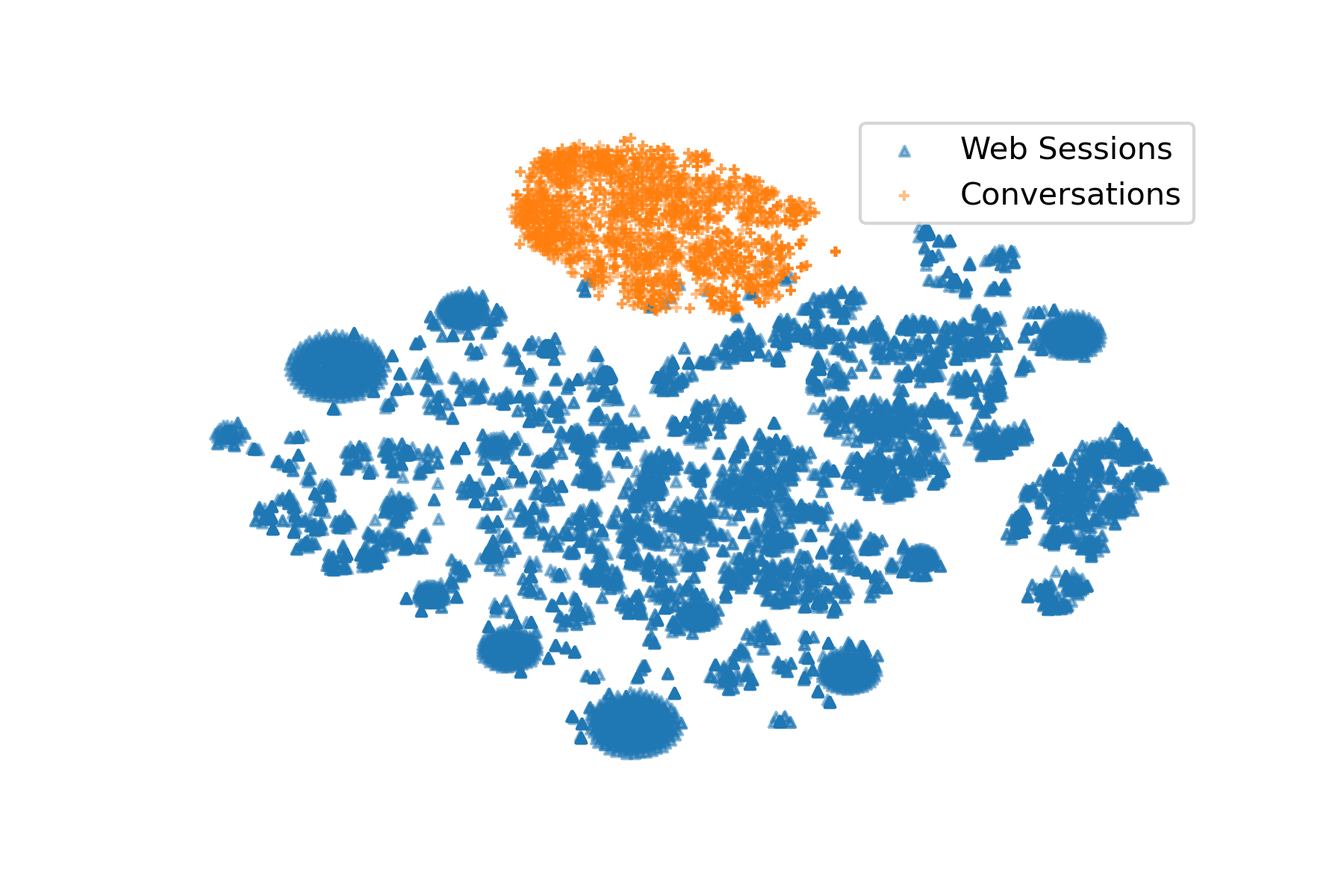}
        \caption{Keyword model}
    \end{subfigure}
    \begin{subfigure}{0.25\textwidth}
        \centering
        \includegraphics[width=\columnwidth]{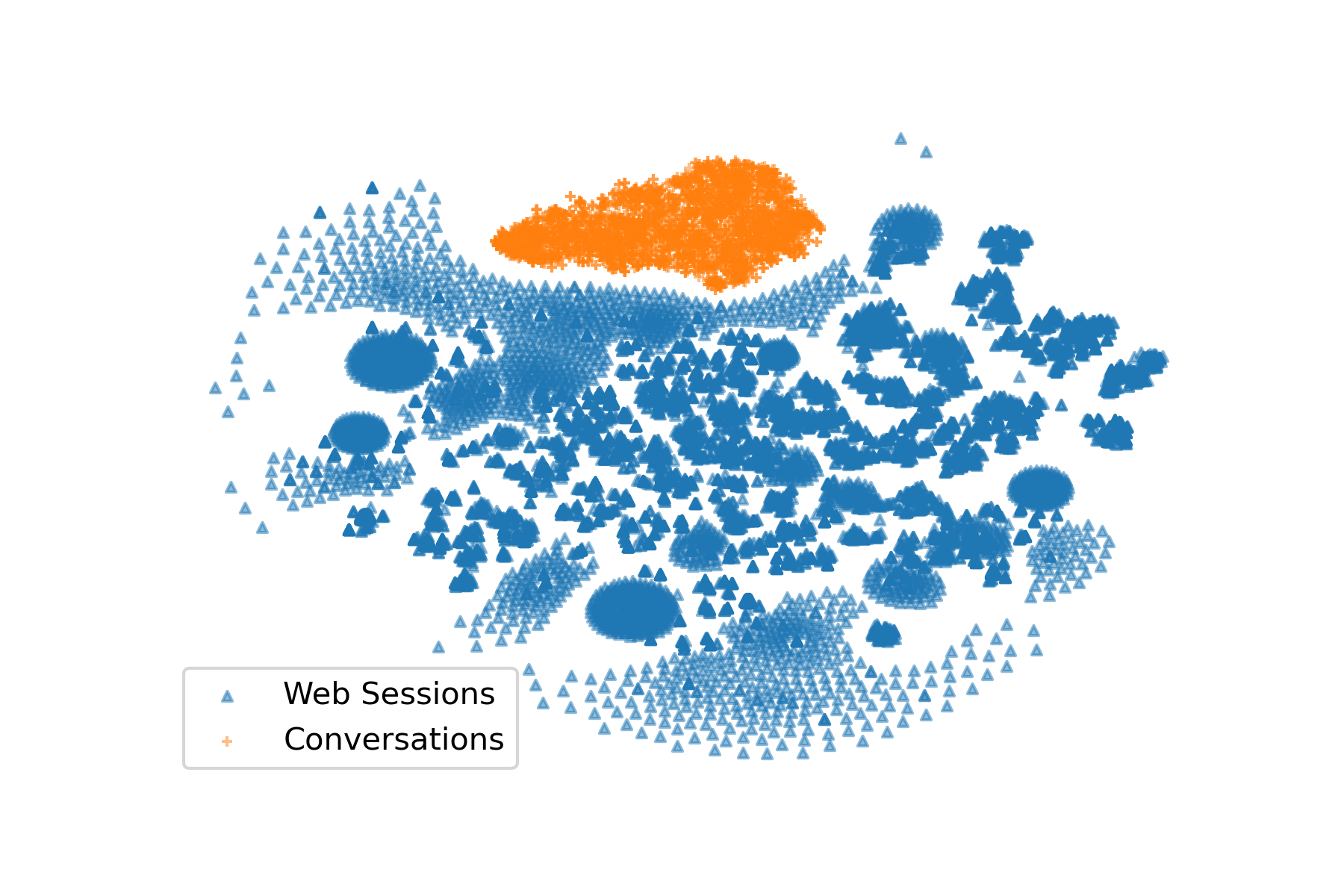}
        \caption{Latent Feature model}
    \end{subfigure}
    \begin{subfigure}{0.25\textwidth}
        \centering
        \includegraphics[width=\columnwidth]{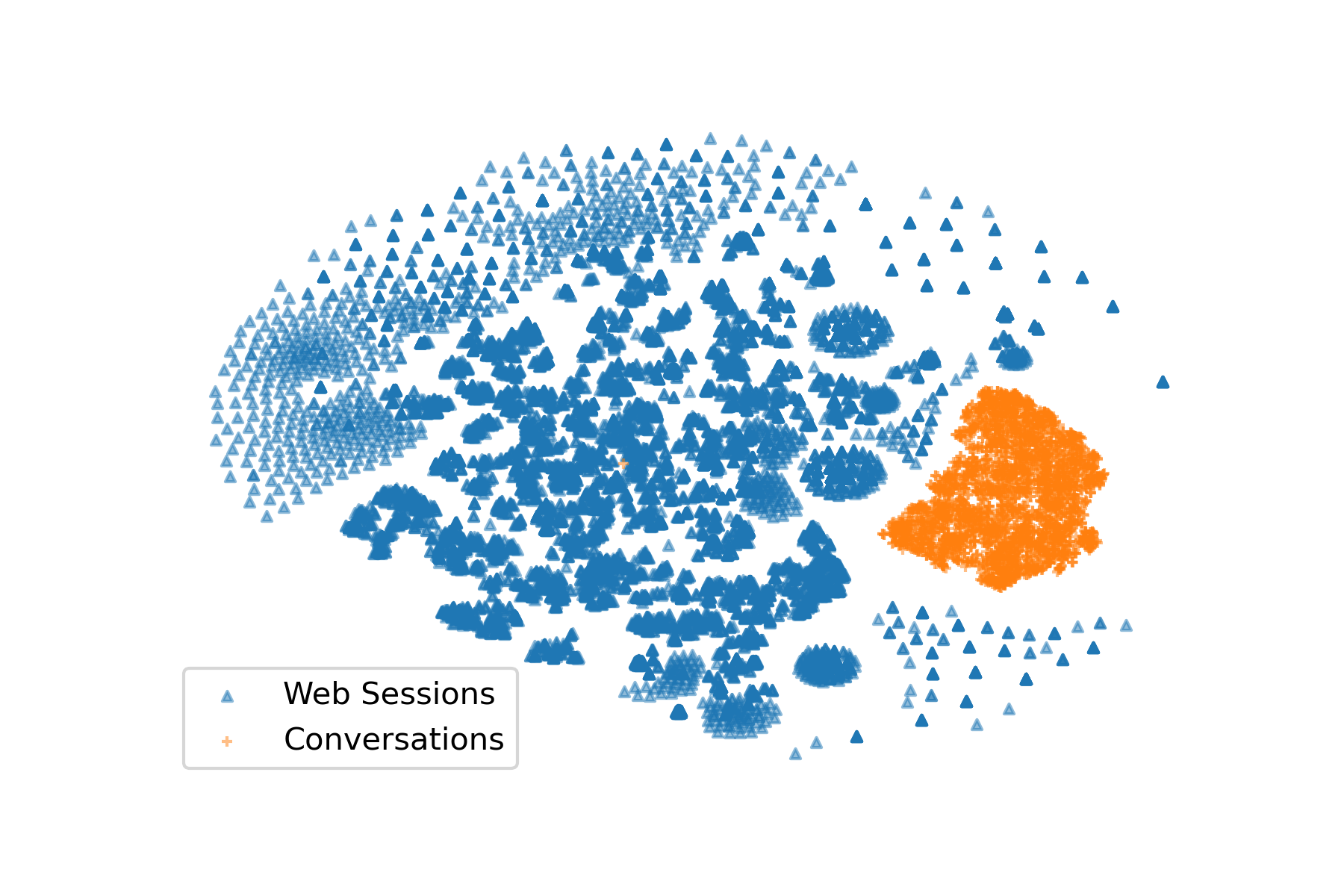}
        \caption{Relative Representation model}
    \end{subfigure}
    \vspace*{-0.5\baselineskip}
    \caption{t-SNE visualization of the input representations.}
    \label{fig:tsne_input}
\end{figure*}

\begin{figure*}[tb]
    \vspace*{-0.5\baselineskip}
    \centering
    \begin{subfigure}{0.25\textwidth}
        \centering
        \includegraphics[width=\columnwidth]{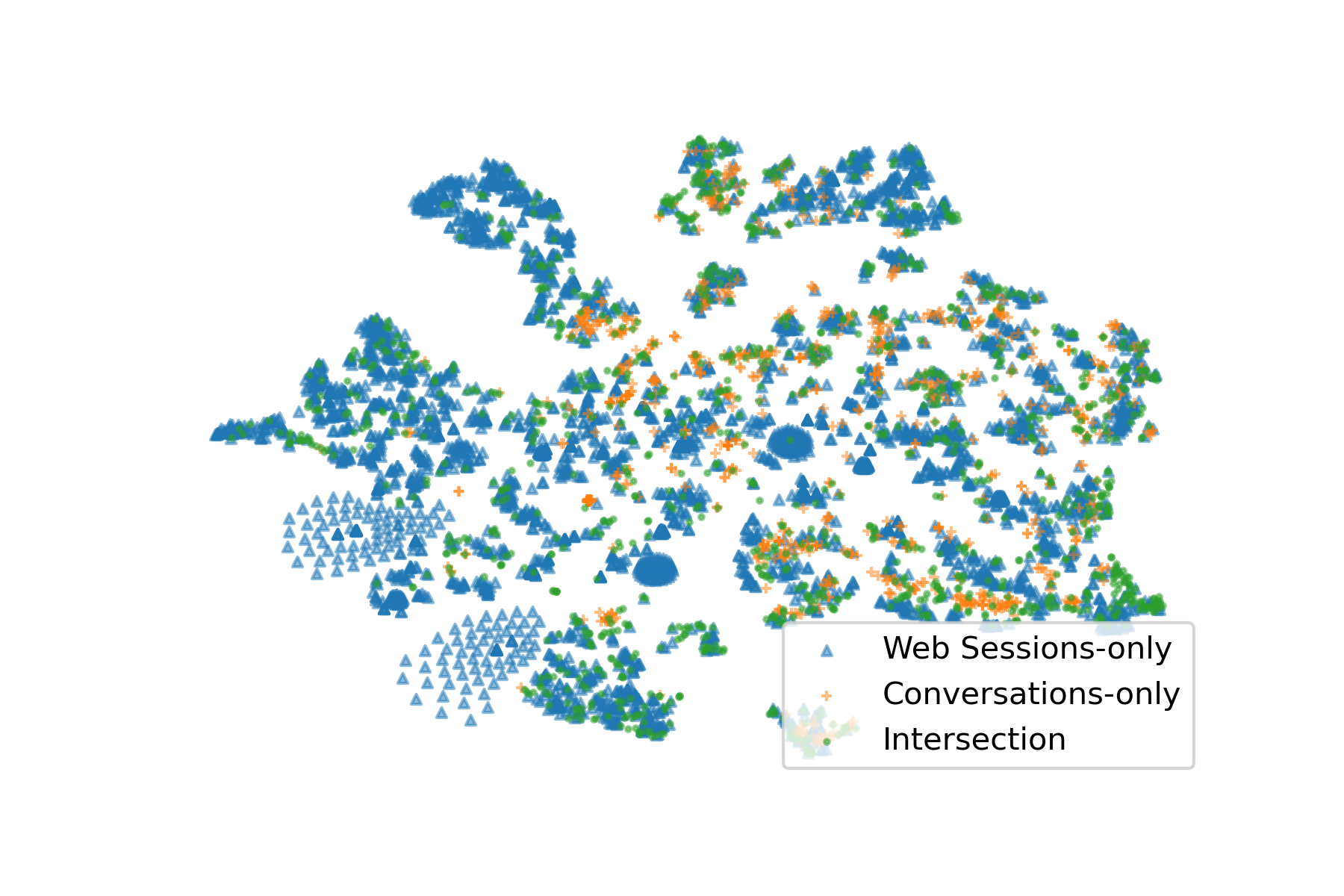}
        \caption{Keyword model}
    \end{subfigure}
    \begin{subfigure}{0.25\textwidth}
        \centering
        \includegraphics[width=\columnwidth]{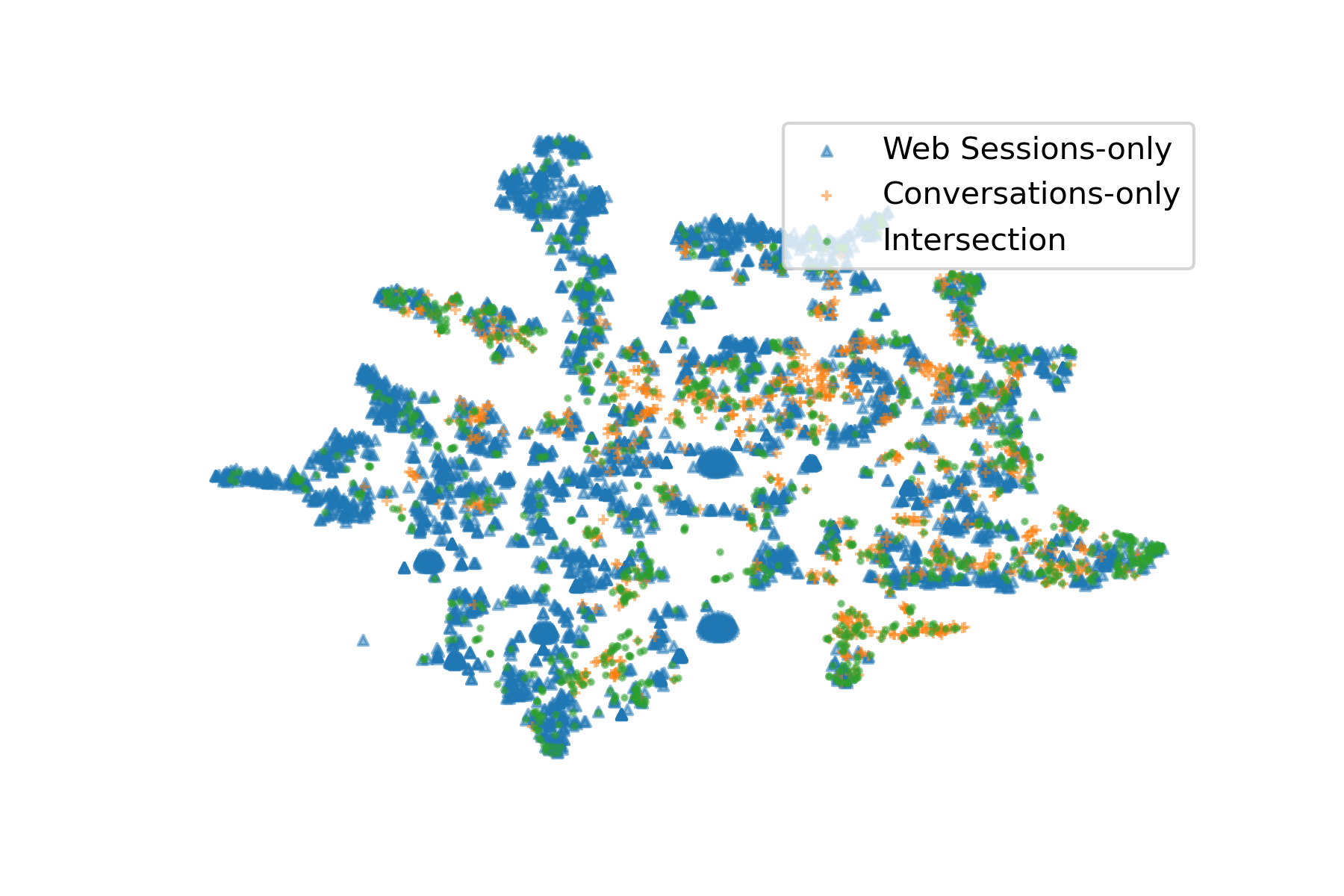}
        \caption{Latent Feature model}
    \end{subfigure}
    \begin{subfigure}{0.25\textwidth}
        \centering
        \includegraphics[width=\columnwidth]{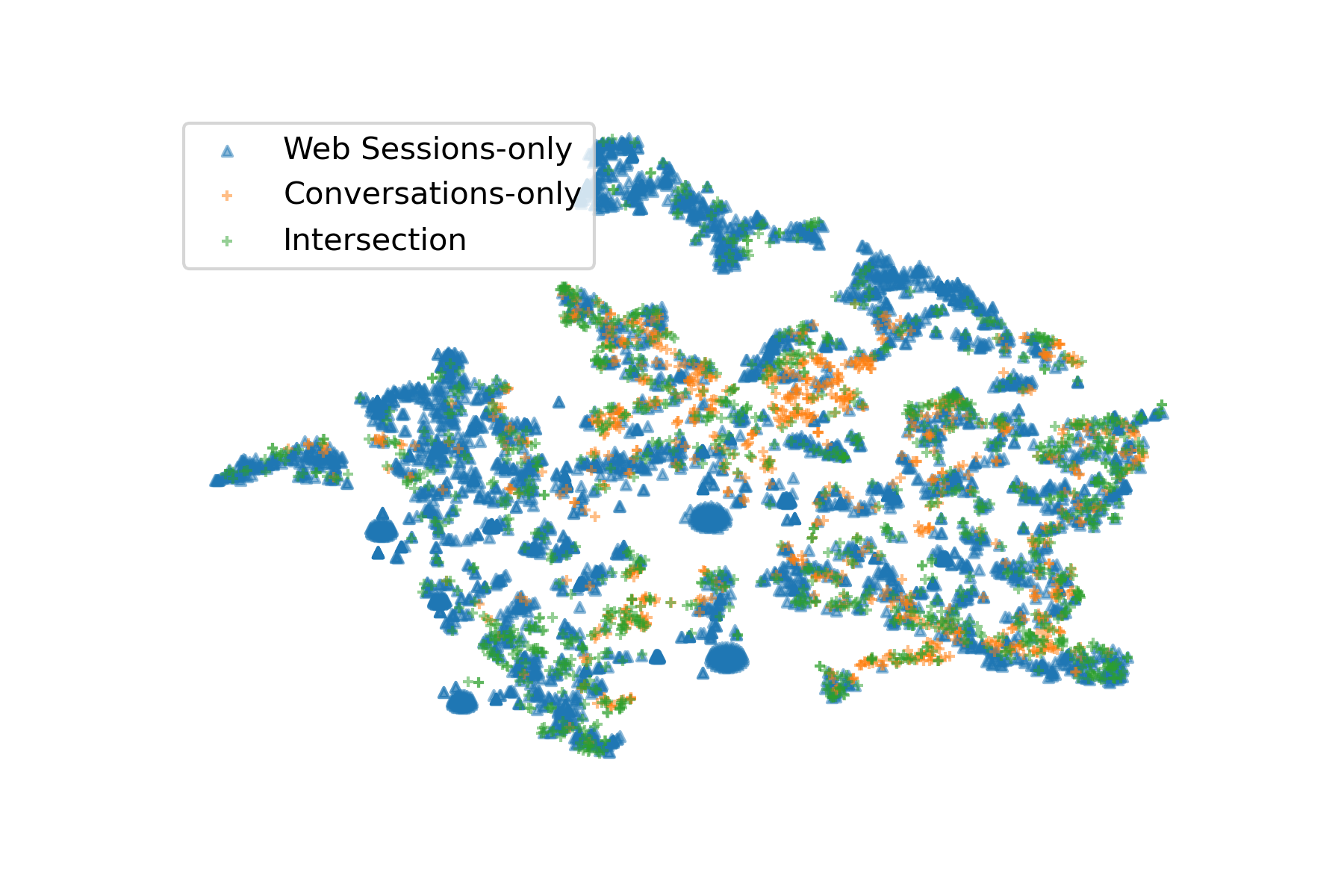}
        \caption{Relative Representation model}
    \end{subfigure}
    \vspace*{-0.5\baselineskip}
    \caption{t-SNE visualization of the model output.}
    \label{fig:tsne_output}
\end{figure*}

We use t-SNE to visualize the input representations of our three different approaches in Fig.~\ref{fig:tsne_input}. That is, we visualize the keywords, the latent features, and the relative representations, respectively. We observe that the conversations and web sessions are clustered together, showing that the two modalities contain different information about the users. It likely differs from RSs for items with multi-modal representations, where for instance an image and a text description of the same item typically contain more similar information about the item. In Fig.~\ref{fig:tsne_output}, we visualize the output after the conversations and web sessions have been passed through the neural models. Now modalities of the same type are more separated, as the different information is learned as signals for the same recommendations. Note that the users with both conversations and web sessions (the intersection) have one joint output.
Overall, the visualization shows that the gain of learning recommendations from multi-modal user interactions is partly because the modalities contain different information about users that complement each other well for the task.

\section{Conclusion and Future Work}
\label{sec:concl}
We have taken an important first step in the problem of learning recommendations from multi-modal user interactions. This is a highly relevant problem to which no satisfying solution currently exists, as prior work and public datasets focused only on items with multi-modal representations and complete modalities.
Our contributed dataset contains real-world user interactions of multiple modalities in terms of website actions and call center conversations that are naturally missing for some users.
Experimental comparison of several approaches for combining the modalities reveals that they contain very different information 
that complement each other well for the recommendation task.
Investigation of three new ways of representing the modalities shows that a model which automatically learns latent features is most effective while a model based on relative representations has the advantage of being less dependent on long user history. Finally, a method using keyword extraction is particularly good at capturing feature interactions between the modalities.
Overall, we demonstrate that it is particularly beneficial to include user interactions of different modalities for generating effective personalized recommendations.


Since this work focuses on the fusion part of different modalities, the conversations are treated as generic text and could be any text related to a user such as e-mails, chats, and social media posts. As future work, we plan to do a dedicated work on effectively learning recommendations from past conversations by taking into account the context, like time and speaker.


\bibliographystyle{ACM-Reference-Format}
\balance
\bibliography{sample-base}


\end{document}